\documentclass[%
 reprint,
 superscriptaddress,
 amsmath,amssymb,
 aps,
footinbib
]{revtex4-2}
\usepackage{graphicx}
\usepackage{subcaption}

\usepackage{bm}
\usepackage[colorlinks=true,linkcolor=blue,urlcolor=black,citecolor=blue,bookmarksopen=true]{hyperref}
\usepackage{amsmath,physics}
\usepackage{xcolor}

\usepackage{cancel}
\begin{document}

\preprint{APS/123-QED}

\title{On the generalized eigenvalue problem in subspace-based excited state methods for quantum computers} 
\author{Prince Frederick Kwao} 
\affiliation{Department of Chemistry, University of North Dakota, ND 58202, USA}
\author{Srivathsan Poyyapakkam Sundar}
\affiliation{Department of Chemistry, University of North Dakota, ND 58202, USA}
\author{Brajesh Gupt}
\affiliation{Department of Chemistry, University of North Dakota, ND 58202, USA}
\author{Ayush Asthana}
\email{ayush.asthana@und.edu}
\affiliation{Department of Chemistry, University of North Dakota, ND 58202, USA}
\begin{abstract}
Solving challenging problems in quantum chemistry is one of the most promising applications of quantum computers. Within the quantum algorithms proposed for problems in excited state quantum chemistry, subspace-based quantum algorithms, including quantum subspace expansion (QSE), quantum equation of motion (qEOM) and quantum self-consistent equation-of-motion (q-sc-EOM), are promising for pre-fault-tolerant quantum devices.
The working equation of QSE and qEOM requires solving a generalized eigenvalue equation with associated matrix elements measured on a quantum computer. 
Our careful analytical and numerical analysis of the standard and generalized eigenvalue problems, especially in the context of excited-state methods, shows that the errors in eigenvalues magnify drastically with an increase in the condition number of the overlap matrix when a generalized eigenvalue equation is solved in the presence of statistical sampling errors. 
This makes methods such as QSE unstable to errors that are unavoidable when using quantum computers. 
Further, at very high condition numbers of the overlap matrix, the QSE's working equation could not be solved without any additional steps in the presence of sampling errors, as it becomes ill-conditioned. It was possible to use the thresholding technique in this case to solve the equation, but the solutions achieved had missing excited states, which may be a problem for future chemical studies. 
We also show that excited-state methods that have an eigenvalue equation as the working equation, such as q-sc-EOM, do not have the problems associated with the condition number and could be generally more stable to errors, and therefore, more suitable candidates for excited-state quantum chemistry calculations using quantum computers. 
\end{abstract}

\maketitle

\section{Introduction}

Quantum computing is expected to greatly enhance the simulation of quantum systems, particularly challenging problems in ground- and excited-state quantum chemistry~\cite{tilly2021variational,mcardle2020quantum,cerezo2021variational,peruzzo2014variational,magann2021pulses,kandala2017hardware,cao2019quantum,fedorov2022vqe,magann2021pulses}. Recent years have seen rapid advances in methods and algorithms for computing the eigenstates of a molecular Hamiltonian in quantum chemistry using quantum computers ~\cite{grimsley2019adaptive, asthana2022equation, kumar2022quantum, bauer2020quantum, stair2021simulating, stair2020multireference,kowalski2020quantum, huggins2020non, lee2018generalized,asthana2022minimizing,meitei2021gate}.
Excited state quantum chemistry is exciting for quantum computing applications because classical computer-based quantum chemistry methods face challenges in several cases where an accurate description of excited states is needed, and quantum computers may provide an advance in these problems.
For instance, leading scalable excited state methods face challenges in the accurate description of doubly excited states~\cite{boguslawski2018targeting,rishi2023dark,ravi2022intermediate}, accurate potential energy surfaces for the study of conical intersections and excited state dynamics~\cite{kohn2007can,thomas2021complex,kjonstad2017crossing,tajti2018accuracy}, especially when the ground state has a multi-reference character~\cite{lischka2018multireference, evangelista2018perspective}. 

Two classes of quantum algorithms have been proposed to compute the potential energy surfaces of excited states: state-specific methods and subspace diagonalization-based methods. 
State-specific methods generally find one or a few excited states at a time by an optimization process similar to the ground state. Several methods in this type are proposed based on quantum phase estimation~\cite{kitaev1997quantum,nielsen2002quantum,russo2021evaluating,bauman2020toward,sugisaki2021quantum,santagati2018witnessing,o2019quantum,sugisaki2021bayesian,sugisaki2022bayesian,poulin2018quantum},   quantum annealing~\cite{teplukhin2021computing,sugisaki2022adiabatic} and variational quantum eigensolver (VQE)~\cite{chan2021molecular,yordanov2021molecular,xie2022orthogonal,cadi2024folded,shirai2022calculation,wen2021variational,mondal2023ground,nykanen2024delta,tilly2020computation,nakanishi2019subspace,wang2023electronic,benavides2024quantum,wang2024quantum}. 
Although some of these methods may provide the exact excited state energies, they may be limited in their predictive power in near-degenerate excited states on imperfect quantum computers.
For pre-fault-tolerant quantum computers, subspace diagonalization-based methods are promising (for a review of subspace methods for quantum computing applications, please refer to Motta et al.~\cite{motta2024subspace}). They provide a range of excited states by diagonalizing the Hamiltonian in a subspace created by electronic excitations on the ground state wavefunction.  
In these methods, a generalized eigenvalue equation or an eigenvalue equation is solved on a classical computer while the elements of these matrices are measured on a quantum computer. Leading examples of methods of this type are Quantum Subspace Expansion (QSE)~\cite{mcclean2017hybrid,urbanek2020chemistry}, Quantum Equation of Motion (qEOM)~\cite{pauline_qeom_2020}, and Quantum Self-consistent Equation-of-Motion (q-sc-EOM)~\cite{asthana2022equation,kumar2023quantum}.
Several approaches for including extended orbital space in subspace-based excited state methods have also been developed~\cite{kumar2022quantum,jensen2024quantum,reinholdt2024subspace,von2024reduced,bauman2024excited,bauman2023coupled}.
Recently developed MORE-ADAPT by Grimsley et al. ~\cite{grimsley2025challenging} is also a related method, which optimizes an adaptively built circuit for a weighted cost function with chosen initial states to provide accurate excited states.
It requires solving an eigenvalue equation but also needs increased quantum resources.
Another promising recent development in subspace methods is the sampled-based quantum diagonalization (SQD)~\cite{barison2024quantum,shajan2024towards}, which diagonalizes Hamiltonian in a subspace built using samples from a quantum state, requires solving an eigenvalue equation on a classical computer similar to q-sc-EOM.

We have previously shown that q-sc-EOM has major theoretical and practical benefits compared with QSE and qEOM~\cite{asthana2022equation}. q-sc-EOM is size-intensive, unlike QSE and satisfies killer condition, unlike qEOM.
The q-sc-EOM method, unlike QSE and qEOM, can be combined with the Davidson algorithm developed by Kim et al.~\cite{kim2023two} to reduce shot requirements by a factor of $O(N^4)$.
Generally applicable error mitigation strategies for the associated quantum linear response formalism were introduced in Ref.~\cite{ziems2025understanding}.
Our analytical and numerical tests have also shown that q-sc-EOM may be more robust to errors in measurement compared with QSE and qEOM~\cite{asthana2022equation}. This is due to the fact that the generalized eigenvalue problem, that is the working equation of QSE and qEOM, becomes sensitive to errors, even becoming singular when the condition number of the overlap matrix becomes too large. 
Singularities were reported in practical calculations using QSE by Barrison et al. in Ref.~\cite{barison2022quantum}. 
Singularities were also reported in the related quantum linear response formalisms that make use of generalized eigenvalue equations by Kjellgren et al. in Ref.~\cite{kjellgren2024divergences}. They argued that the choice of orbitals becomes important in such cases. 
This problem is also seen in other ground state quantum algorithms which have similar generalized eigenvalue equations as the working equation, such as quantum Krylov-based methods~\cite{cortes2022quantum,yu2025quantum,anderson2024solving,tkachenko2021correlation,parrish2019quantum}, non-orthogonal quantum eigensolver~\cite{huggins2020non,baek2023say}, connected moment expansion~\cite{zheng2024unleashed,kowalski2020quantum}, etc.
Epperly et al. carried out a detailed study that lays down the mathematical foundations of such subspace diagonalization~\cite{epperly2022theory}. They show that the thresholding technique, which works by deleting small valued eigenvalues and associated eigenvectors of the overlap matrix and re-projecting the Hamiltonian on the resulting space, is effective at accurately solving for the lowest eigenvalue in generalized eigenvalue problems in case of singularities. 

 In this paper, we report both analytical and numerical analyses of excited-state methods, QSE and q-sc-EOM, that require solving a generalized eigenvalue equation and an eigenvalue equation, respectively, in the presence of statistical sampling errors. 
Our analysis results in the following contributions in this paper: (a) we translate perturbation bounds into practical diagnostics linking $\lambda_{\min}(\mathbf{S})$, expected shot noise, and excited-state reliability, (b) show numerically that ‘moderate’ condition numbers can already induce large instability under shot noise in QSE (H4 and NH3 cases), (c) demonstrate that thresholding can stabilize solvability but can reduce spectral completeness (missing excited states) in the studied examples, (d) contrast with an orthonormal-overlap formulation (q‑sc‑EOM) to isolate the role of generalized eigenproblem conditioning.

  
The sensitivity of generalized eigenvalue problems to ill-conditioning is a classical result in numerical linear algebra and has been recognized in prior quantum subspace diagonalization work (including Ref. ~\cite{epperly2022theory}). Our contribution is not a new perturbation theorem; rather, we quantify and illustrate how this mechanism manifests in realistic excited-state subspace algorithms with unavoidable sampling noise, and we highlight consequences that matter for chemistry use cases (e.g., spectral incompleteness under stabilization). 

\section{Theory}
\subsection{Quantum subspace expansion}
The QSE algorithm involves creating a subspace using the action of excitation operators on the ground state wavefunction ($\ket{\Psi_{gr}}$) prepared on a quantum computer using any quantum algorithm. 
These excitations are used from an excitation manifold given by, $\{\hat{I}\cup\hat{G}_I\}$, where $\hat{G}_I$ includes single and double electronic excitations, given by
\begin{align}
    \begin{split}
    \hat{G}_{i}^{a} &= \hat{a}^{\dagger}_a\hat{a}_i,\\
    \hat{G}_{ij}^{ab} &= \hat{a}^{\dagger}_a\hat{a}^{\dagger}_b\hat{a}_j\hat{a}_i,
    \end{split}
\end{align}
and $\hat{I}$ is the Identity operator. 
Here, $\hat{a}^{\dagger}_a(\hat{a}_a)$ are creation(destruction) operators acting on the orbital $a$.
If $\ket{\Psi_{VQE}}$ is the ground state wavefunction prepared using a VQE algorithm, the subspace can be created by the action of the excitation operators on $\ket{\Psi_{VQE}}$.

Matrix elements of matrices $\mathbf{H}$ and $\mathbf{S}$ are then estimated on a quantum computer for the generalized eigenvalue equation, 
\begin{align*}
    \mathbf{H}C=\mathbf{S}CE,
\end{align*}
where $\mathbf{H}$ and $\mathbf{S}$ are given by
\begin{align*}
    \mathbf{H}_{I,J} &= \bra{\Psi_{\text{VQE}}}\hat{G}^{\dagger}_I\hat{H}\hat{G}_J\ket{\Psi_{\text{VQE}}},\\
    \mathbf{S}_{I,J} &= \bra{\Psi_{\text{VQE}}}\hat{G}^{\dagger}_I\hat{G}_J\ket{\Psi_{\text{VQE}}},
\end{align*}
where $\hat{H}$ is the molecular Hamiltonian. Once the matrices are formed, the generalized eigenvalue problem is solved on a classical computer, resulting in the energy of the ground and excited states of the quantum system as eigenvalues $E_k$.

\subsection{Quantum self-consistent equation of motion}
The q-sc-EOM method is based on the equation of motion theory, which was developed for classical quantum chemistry by Rowe et al.~\cite{rowe1968equations} and first applied to quantum computing by Ollitrault et al.~\cite{pauline_qeom_2020}.
The q-sc-EOM method introduced self-consistent operators, originally developed by Prasad et al.~\cite{prasad1985some}, within the qEOM framework~\cite{asthana2022equation}.
The working equation of the q-sc-EOM method takes a similar form to QSE. It involves solving the eigenvalue equation, given by
\begin{align}\label{qsceombasic}
    \mathbf{H}C=CE,
\end{align}
where elements of $\mathbf{H}$ matrix are given by
\begin{align}
    \mathbf{H}_{I,J}=\bra{\Psi_{\text{HF}}}\hat{G}_I^{\dagger}U^{\dagger}(\theta)\hat{H}U(\theta)\hat{G}_J\ket{\Psi_{\text{HF}}} - \delta_{I,J}E_{\text{HF}}.
\end{align}
Here, $U(\theta)$ is the unitary that represents the quantum circuit in a VQE with the parameters the same as solved for the ground state problem, such that
\begin{align}
    \ket{\Psi_{\text{VQE}}}=U(\theta)\ket{\Psi_{\text{HF}}}
\end{align}
and $\ket{\Psi_{HF}}$ is the Hartree-Fock state. Finally, the simple eigenvalue equation is solved (or the matrix $\mathbf{H}$ is diagonalized) on a classical computer to find the energy eigenvalues of a molecular system.
  In this work, when we highlight stability advantages associated with avoiding a generalized eigenvalue problem, we specifically refer to q‑sc‑EOM, whose overlap matrix is analytically the identity due to the orthonormal construction of the basis. Standard qEOM retains a generalized eigenvalue structure and does not generally share this conditioning immunity.

\subsection{ Eigenvalue problem vs generalized eigenvalue problem}\label{GEP}
A key difference between the two methods is that q-sc-EOM requires solving an eigenvalue problem since the overlap matrix is strictly an identity matrix proved analytically. 
QSE on the other hand, requires solving a generalized eigenvalue equation on a classical computer. 
In the presence of errors in the $\mathbf{H}$ and $\mathbf{S}$ matrix, this step can be challenging. This is because solving a generalized eigenvalue equation requires two steps, where the first step is the conversion of the generalized eigenvalue equation into an eigenvalue equation by modifying the eigenvector matrix as
\begin{align}
    C = XC^\prime, 
\end{align}
where $X = \mathbf{S}^{-\frac{1}{2}}$ and multiplying by $X^\dagger$ on the left. This results in an effective eigenvalue equation
\begin{align}
    \mathbf{H}^\prime C^\prime = C^\prime E,
\end{align}
where $\mathbf{H}^\prime = X^\dagger \mathbf{H}X$.

This process requires finding the inverse of the overlap matrix ($\mathbf{S}^{-\frac{1}{2}}$). The second step in this process is solving the new eigenvalue problem. 
Inverting a matrix in the first step in the presence of errors in the $\mathbf{S}$ matrix can be a challenging step, and its accuracy would depend strongly on the condition number of the overlap matrix and the error rates. Conventional knowledge suggests that if the condition number of the $\mathbf{S}$ matrix is large, it can become unstable.

If the condition number is ``too large'', it can lead to singularities as seen in many other studies on classical and quantum algorithms. Quantum algorithms are especially vulnerable because, while classical computers work at numerical accuracy of $\sim$10$^{-16}$, even perfect quantum computers will always have statistical sampling errors that are inherent in quantum computation. 
In prior studies, many efforts have been made to solve the problem of singularities in generalized eigenvalue equations. 
Several practical methods have been employed for solving such problems, including the thresholding method~\cite{epperly2022theory}, the deflation method~\cite{hemmatiyan2018excited,hemmatiyan2019unraveling,greenman2008electronic,mazziotti2003extraction}, and the shift method~\cite{hemmatiyan2018excited,hemmatiyan2019unraveling,greenman2008electronic,mazziotti2003extraction}, with thresholding as the most common choice for quantum algorithms. 
The thresholding technique involves re-projecting the Hamiltonian in a reduced subspace of eigenvectors of the overlap matrix, with corresponding eigenvalues greater than a threshold. 

\subsection{Error analysis in standard and generalized eigenvalue equation}
In this subsection, we will provide a concise theoretical analysis of errors in eigenvalues in standard and generalized eigenvalue equations using matrix perturbation theory. For detailed analysis, please refer to Ref.~\cite{epperly2022theory}.

\subsubsection{Standard eigenvalue equation}
In our study, the Hamiltonian, $\mathbf{H}$, is Hermitian and the overlap matrix, $\mathbf{S}$, is positive definite. The standard eigenvalue equation is given by
\begin{align}
    \mathbf{H}x=\lambda x,
\end{align}
where $\lambda$ and $x$ are the eigenvalue and eigenvector, respectively.
The perturbed matrix and eigenvalues can be written as
\begin{align}
    \tilde{\mathbf{H}} &= \mathbf{H}+\Delta \mathbf{H},\\
    \tilde{\lambda} &= \lambda + \delta\lambda.
\end{align}
Here, $\Delta \mathbf{H}_{i,j}\leq \epsilon$, with $\epsilon$ being the maximum allowed error (dependent on the number of shots used).
The perturbed eigenvalue equation can be written as
\begin{align}
    (\mathbf{H}+\Delta \mathbf{H})(x+\delta x)=(\lambda + \delta \lambda)(x+\delta x).
\end{align}
Taking the first-order terms in error and rearranging, we get the first-order eigenvalue shift
\begin{align}
    \delta \lambda^{(1)} = x^\dagger \Delta \mathbf{H}\, x.
\end{align}
For any vector $x$, we know that~\cite{horn1994topics}
\begin{align}
    |x^\dagger \Delta \mathbf{H} x| \leq ||\Delta \mathbf{H}||_2 \leq ||\Delta \mathbf{H}||_F,
\end{align}
where $||\cdot||_2$ and $||\cdot||_F$ are the spectral and Frobenius norms, respectively.
Since,
\begin{align}
    ||\Delta \mathbf{H}||^2_F = \sum_{i,j}|\Delta \mathbf{H}_{i,j}|^2 \leq n^2\epsilon^2,\\
    \implies ||\Delta \mathbf{H}||_F\leq n\epsilon,\label{ne}
\end{align}
where $n$ is the dimension of $\mathbf{H}$. We can state that
\begin{align}
    |\delta \lambda^{(1)}|&\leq n\epsilon,\\
    |\tilde{\lambda}-\lambda| &= O(n\epsilon)+O(\epsilon^2).
\end{align}
This means that the error in eigenvalues for the standard eigenvalue equation scales linearly in $\epsilon$ with a worst-case pre-factor of $n$.

\subsubsection{Generalized eigenvalue equation}
We start with the generalized eigenvalue equation
\begin{align}
    \mathbf{H}x=\lambda \mathbf{S}x.
\end{align}
We choose eigenvectors normalized in the $\mathbf{S}$-inner product as $x^\dagger \mathbf{S} x =1$.
The perturbed generalized eigenvalue equation, following a similar process as in the standard eigenvalue equation, can be written as
\begin{align}
    (\mathbf{H}+\Delta \mathbf{H})(x+\delta x)=(\lambda+\delta \lambda)(\mathbf{S}+\Delta \mathbf{S})(x+\delta x).
\end{align}
Collecting first-order terms in error and rearranging, we get
\begin{align}
    \delta \lambda^{(1)}=x^\dagger (\Delta \mathbf{H}-\lambda \Delta \mathbf{S}) x.
\end{align}
Using the operator norm bound, we have
\begin{align}
    |x^\dagger (\Delta \mathbf{H}-\lambda \Delta \mathbf{S}) x|\leq ||\Delta \mathbf{H}-\lambda \Delta \mathbf{S}||_2 \, ||x||_2^2.\label{1}
\end{align}
Since $\mathbf{S}\succ 0$, all eigenvalues of $\mathbf{S}$ are positive, and
\begin{align}
    1=x^\dagger \mathbf{S} x\geq \lambda_{\min}(\mathbf{S})\,||x||_2^2,
\end{align}
giving the relationship
\begin{align}
    ||x||_2^2\leq \frac{1}{\lambda_{\min}(\mathbf{S})}.\label{2}
\end{align}
Also, by the triangle inequality~\cite{horn1994topics} and Eq.~\eqref{ne},
\begin{align}
    ||\Delta \mathbf{H}-\lambda \Delta \mathbf{S}||_2 \leq ||\Delta \mathbf{H}||_2+|\lambda|\,||\Delta \mathbf{S}||_2 \nonumber\\
    \leq ||\Delta \mathbf{H}||_F+|\lambda|\,||\Delta \mathbf{S}||_F \leq n\epsilon+|\lambda|\,n\epsilon = n(1+|\lambda|)\epsilon.\label{3}
\end{align}
Finally, using Eqs.~\eqref{1}, \eqref{2} and \eqref{3}, we get
\begin{align}
    |\delta \lambda|&\leq \frac{n(1+|\lambda|)}{\lambda_{\min}(\mathbf{S})}\epsilon,\\
    |\tilde{\lambda}-\lambda| &= O\Big(\frac{n(1+|\lambda|)}{\lambda_{\min}(\mathbf{S})}\epsilon\Big)+O(\epsilon^2).
\end{align}
This means that the error in eigenvalues for the generalized
eigenvalue equation still scales linearly in $\epsilon$ but with a worst-case
pre-factor of $n\frac{(1+|\lambda|)}{\lambda_{\min}(\mathbf{S})}$.
Thus, the errors magnify in case of small $\lambda_{\min}(\mathbf{S})$ or large $\lambda$.

\subsubsection{Practical shot-noise guidance}\label{practical}
The conservative perturbation bound derived above can be converted into a practical guideline in the presence of statistical sampling noise (shot noise). 
If each matrix element entering $\mathbf{H}$ and $\mathbf{S}$ is estimated using $N_{\text{shots}}$ measurements, then the typical statistical estimation error scales as
\begin{align}
    \epsilon \approx \frac{\sigma}{\sqrt{N_{\text{shots}}}},
\end{align}
where $\sigma$ is an observable- and estimator-dependent constant set by the measurement variance (often $\sigma=O(1)$ for bounded observables). 
Substituting this scaling into the generalized-eigenvalue bound gives the worst-case estimate
\begin{align}
    |\delta \lambda| \lesssim \frac{n(1+|\lambda|)}{\lambda_{\min}(\mathbf{S})}\,\frac{\sigma}{\sqrt{N_{\text{shots}}}}.
\end{align}
Therefore, to achieve a target eigenvalue accuracy $|\delta\lambda|\le \delta$, the required number of shots per matrix element scales as
\begin{align}
    N_{\text{shots}} \gtrsim 
    \left(\frac{\sigma\,n(1+|\lambda|)}{\lambda_{\min}(\mathbf{S})\,\delta}\right)^2.
\end{align}
Equivalently, for a fixed shot budget $N_{\text{shots}}$ and tolerance $\delta$, a necessary (worst-case) condition for reliability is
\begin{align}
    \lambda_{\min}(\mathbf{S}) \gtrsim \frac{\sigma\,n(1+|\lambda|)}{\delta\sqrt{N_{\text{shots}}}}.
\end{align}
These relations make explicit that the sampling cost grows quadratically as $\lambda_{\min}(\mathbf{S})$ decreases (and, in typical cases where $\lambda_{\max}(\mathbf{S})=O(1)$, grows approximately as $\kappa(\mathbf{S})^2$). 
This scaling is qualitatively consistent with the rapid growth of the mean error and variance in QSE as the overlap-matrix condition number increases in the medium-condition-number regime (Fig.~2). 
Moreover, for the case with $\kappa(\mathbf{S})=592.5$, the substantially larger shot requirements observed for QSE relative to q-sc-EOM to achieve comparable accuracy (Fig.~5) are in line with the presence of the amplification factor $1/\lambda_{\min}(\mathbf{S})$ in the generalized-eigenvalue bound.

This diagnostic also suggests when stabilization becomes necessary: when $\lambda_{\min}(\mathbf{S})$ becomes comparable to (or smaller than) the effective noise floor set by $\epsilon$, the generalized-eigenvalue solve can become unreliable, motivating procedures such as thresholding. 
However, such stabilization effectively reduces the working subspace and may compromise spectral completeness, consistent with the missing states observed when thresholding is required in the high-condition-number regime (Fig.~\ref{fig:threshold}). 
Finally, the reduction in QSE errors observed when the overlap matrix is made artificially exact (Fig.~\ref{fig:qseexact}) supports the interpretation that noise in $\mathbf{S}$, and its subsequent inversion in the generalized-eigenvalue solve, is a key driver of the observed instability.
In contrast, for formulations with an orthonormal basis where $\mathbf{S}=\mathbf{I}$ (e.g., q-sc-EOM), $\lambda_{\min}(\mathbf{S})=1$ and this particular amplification mechanism is absent, consistent with the stability observed across the corresponding examples (e.g., Fig.~\ref{fig:592qsceom}).

\begin{figure*}[t]
    \centering
    
\begin{subfigure}[b]{0.45\textwidth}
    \includegraphics[width=\textwidth]{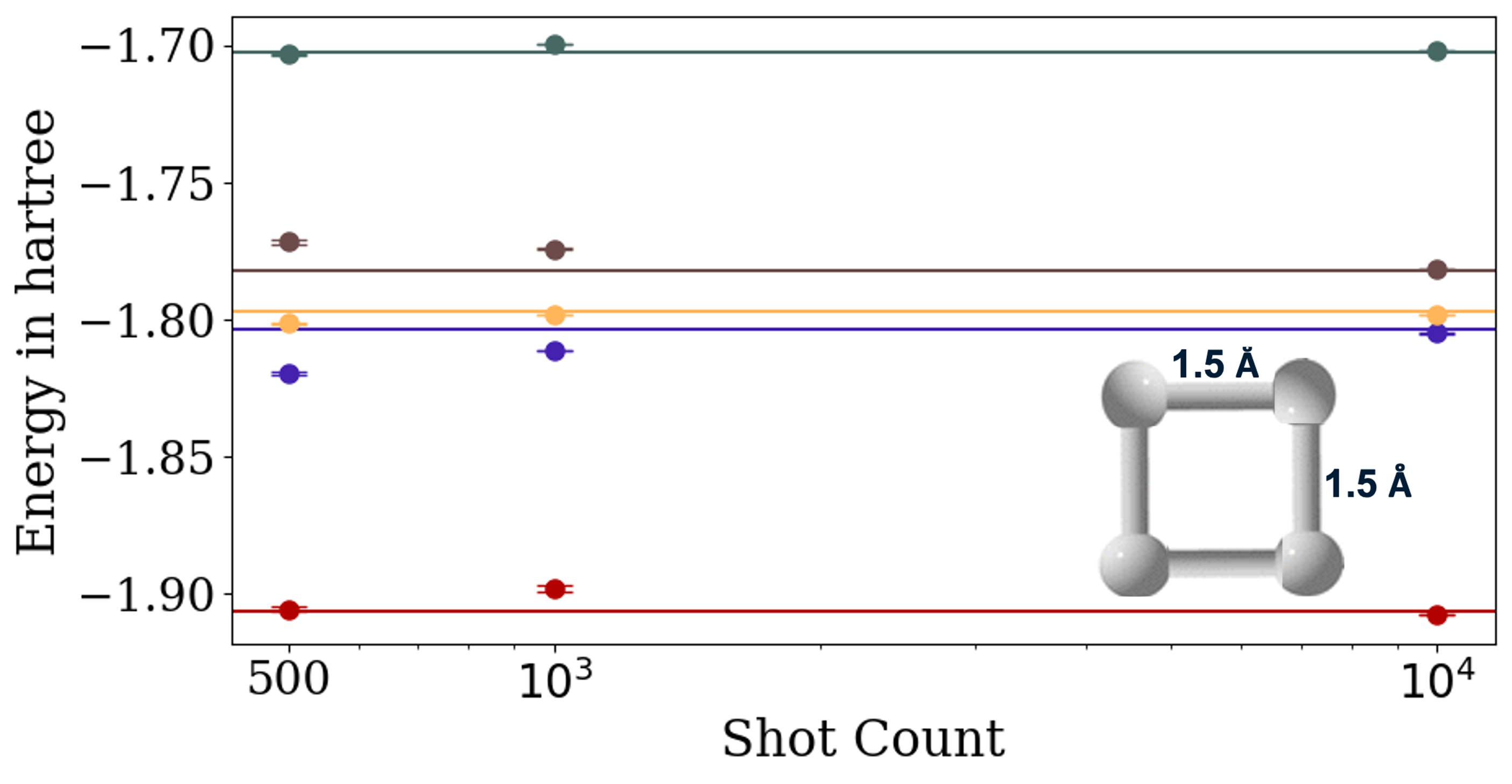}
    \caption{QSE}
\end{subfigure}
\begin{subfigure}[b]{0.45\textwidth}
    \includegraphics[width=\textwidth]{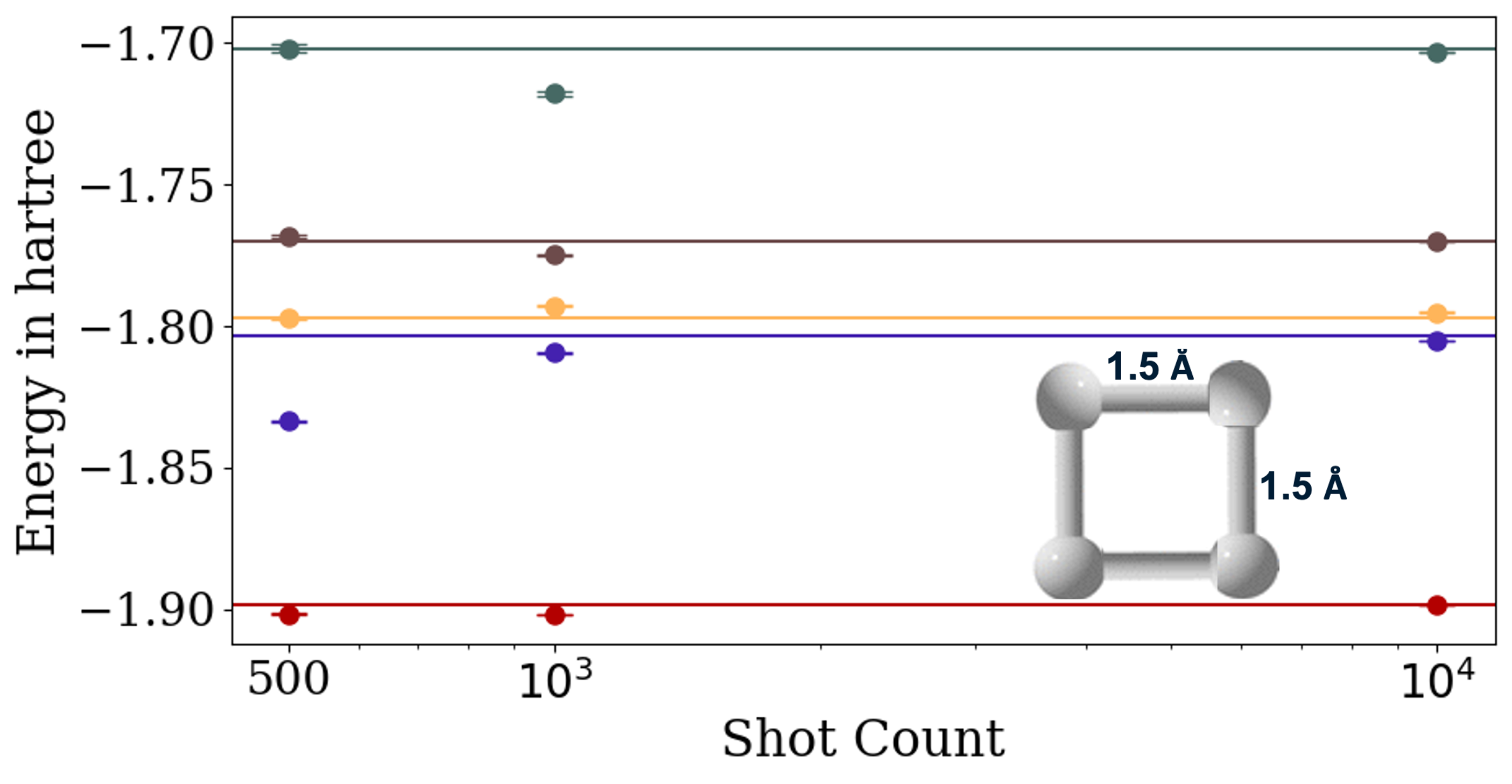} 
    \caption{q-sc-EOM}
\end{subfigure}

\caption{Eigenstates computed using (a) QSE in the case of the low condition number of the overlap matrix (condition number is 7.1) and (b) q-sc-EOM. The solid lines represent the exact solution (corresponding to an infinite number of shots) while the spheres and bars represent the mean and variance of the data in 10 simulations using the number of shots per matrix element on the \textit{x}-axis. The geometry of the H$_4$ molecule is shown in the figure.  This figure illustrates that when the overlap matrix is well-conditioned ($\kappa(\mathbf{S})=7.1$), QSE and q-sc-EOM show comparable shot-noise convergence and variance, i.e., overlap-conditioning-driven error amplification is not dominant in this regime.}
\label{fig:pes.png}
\end{figure*}

\subsection{Computational cost}
The cost of subspace excited state methods can be discussed in circuit depth and the additional number of shots needed after the estimation of the ground state (summarized in table \ref{cost}). 
In terms of the circuit depth, there are no additional gates needed in the circuit in QSE and qEOM beyond the ground state circuit and a negligible number of additional gates are needed (up to 7 CNOT gates) in q-sc-EOM. 
A major limitation of subspace excited state methods, including qEOM, QSE and brute force q-sc-EOM, is the shot count scaling for measuring all the relevant matrix elements in the Hamiltonian ($\mathbf{H}$) and overlap matrices ($\mathbf{S}$). 
The additional shot count scaling in QSE, qEOM and brute force q-sc-EOM are all $O(N^{12})$, where $N$ is the number of orbitals, due to the numerous matrix elements. A major advantage in q-sc-EOM in this regard is that the quantum version of the Davidson algorithm, developed by Kim and Krylov in Ref. \cite{kim2023two}, brings down the cost dramatically to $O(N^{8})$ without any approximation at the cost of additional circuit depth. Additional methods could further reduce the measurement cost in the Hamiltonian expectation value estimation step, thereby lowering the overall resource requirements.

\begin{table}[t]
\centering
\caption{Computational cost in terms of additional circuit depth (in terms of scaling of CNOT gates) and additional shot count scaling of the three subspace excited state methods discussed. $N$ is the generic variable for the number of orbitals. }
\begin{tabular}{|l|c|c|}
\hline
\textbf{Method} &  \textbf{Shot count scaling} \\
\hline
QSE &  $O(N^{12})$ \\
qEOM & $O(N^{12})$ \\
q-sc-EOM & $O(N^{12})$ \\
q-sc-EOM with Davidson  & $O(N^{8})$ \\
\hline
\end{tabular}

\label{cost}
\end{table}

\section{Method}

All computations in this study are carried out in the STO-3G basis using Jordan--Wigner mapping to represent orbitals on qubits. All tests are carried out on an 8-qubit problem for the H$_4$ molecule with different bond lengths to enable a consistent analysis throughout this study.
 The UCCSD ansatz within the VQE algorithm was used in all calculations for the estimation of an accurate ground state, which was followed by QSE or q-sc-EOM methods. The ground-state energy was computed using an infinite number of shots (exact estimation) to ensure all the effects studied resulted from the excited state method. The initial state was taken as the Hartree-Fock state. 
 The parameters of UCCSD were optimized in the exact simulator using the L-BFGS optimizer to reach the noiseless UCCSD ground state.
 We then proceeded to the excited-state calculation using QSE and q-sc-EOM.  
 
 The q-sc-EOM and QSE excited-state calculations are carried out with shot noise (sampling errors) included, corresponding to a fixed number of shots for each study.  
 
 PennyLane~\cite{bergholm2018pennylane}, PySCF~\cite{pyscf}, and Qiskit~\cite{aleksandrowicz2019qiskit} were used to generate the codes for all calculations, which are provided in Refs.~\cite{gitqse,gitqsceom}. The codes were verified using an independent implementation in the PennyLane software and an implementation in the ADAPT-VQE code~\cite{giteom}.

\section{Results and Discussion}
In the following subsections, we will provide a comparative analysis between QSE and q-sc-EOM in various situations of the condition number of the overlap matrix in QSE. 
It should be noted that although we are only using the cases of QSE and q-sc-EOM, the general conclusions are valid for any subspace-based excited state method that relies on solving a generalized eigenvalue equation (or eigenvalue equation) with elements measured on a quantum computer. 
Although there are no strict definitions, we will use the term ``low condition number'' for any matrix condition number low enough that it has almost no effect on the resulting eigenvalues. 
In the H$_4$ cases studied, we have classified any condition number below 10$^2$ in this range. 
 We define ``medium condition number'' as the range in which the condition number is not small enough so that its effects on the results can be ignored, but not large enough that techniques such as thresholding can be used. In the case of an 8-qubit H$_4$ problem, we believe that this range is roughly between 10$^2$ to 10$^4$.
 A ``large condition number'' is defined in cases where the generalized eigenvalue is significantly ill-conditioned and cannot be solved without the use of the thresholding technique. In the H$_4$ cases studied, we have classified any condition number above 10$^4$ into this class. These numerical ranges are system- and noise-level dependent; a more general diagnostic is provided in Section~\ref{practical} in terms of $\lambda_{\min}(\mathbf{S})$ relative to the expected sampling error.
These ranges may change with the size of the matrices involved, but the general conclusions in this study should hold for systems of any size.

\subsection{Low condition number case}

Based on the numerical experiments in our study, we find that QSE and q-sc-EOM have similar resilience to statistical sampling errors in cases where the condition number of the overlap matrix in QSE is low. An example is discussed in Fig. \ref{fig:pes.png}, which presents the energy of the five lowest excited states of a square H$_{4}$ molecular system at a bond distance of 1.5~$\text{\AA}$ computed using QSE and q-sc-EOM. We note that the condition number of the overlap matrix in QSE in this example is a low number of 7.1.
The colored solid lines represent the computed estimates of results at infinite shots, while the dots with bars represent the mean and variance of the computed eigenenergy at the finite number of shots that are indicated on the \textit{x}-axis.
The plot shows the influence of statistical sampling errors on the accuracy of energy measurements.
As the shot count increases, the measured energy values gradually converge towards the reference energies, and it takes about 10,000 shots per matrix element in QSE and q-sc-EOM to reach 0.1 eV.
 Additionally, as expected, the variance in energy values decreases as the measurement counts used increase. 
 The above-mentioned statistical behavior remains consistent with both QSE and q-sc-EOM, and we generally see similar results from both methods in case the condition number of the overlap matrix remains low, as discussed in this example. 
  
\begin{figure*}[t]
    \centering
    
    \begin{subfigure}[b]{0.45\textwidth}
        \centering
        \includegraphics[width=\textwidth]{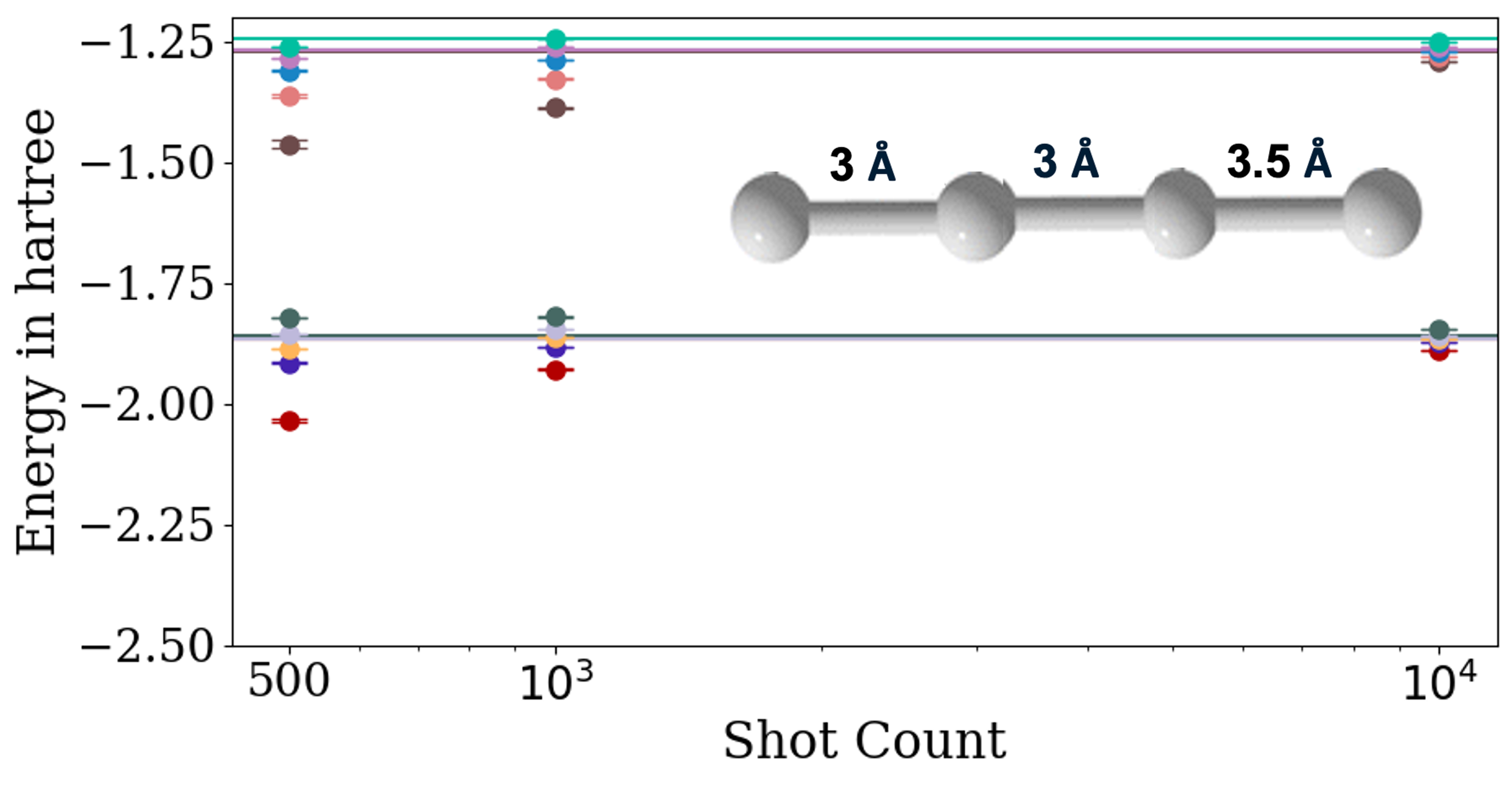}  
        \caption{condition number of 121.5}
        \label{fig:fig1}
    \end{subfigure}
    \hspace{0.5cm}  
    \begin{subfigure}[b]{0.45\textwidth}
        \centering
        \includegraphics[width=\textwidth]{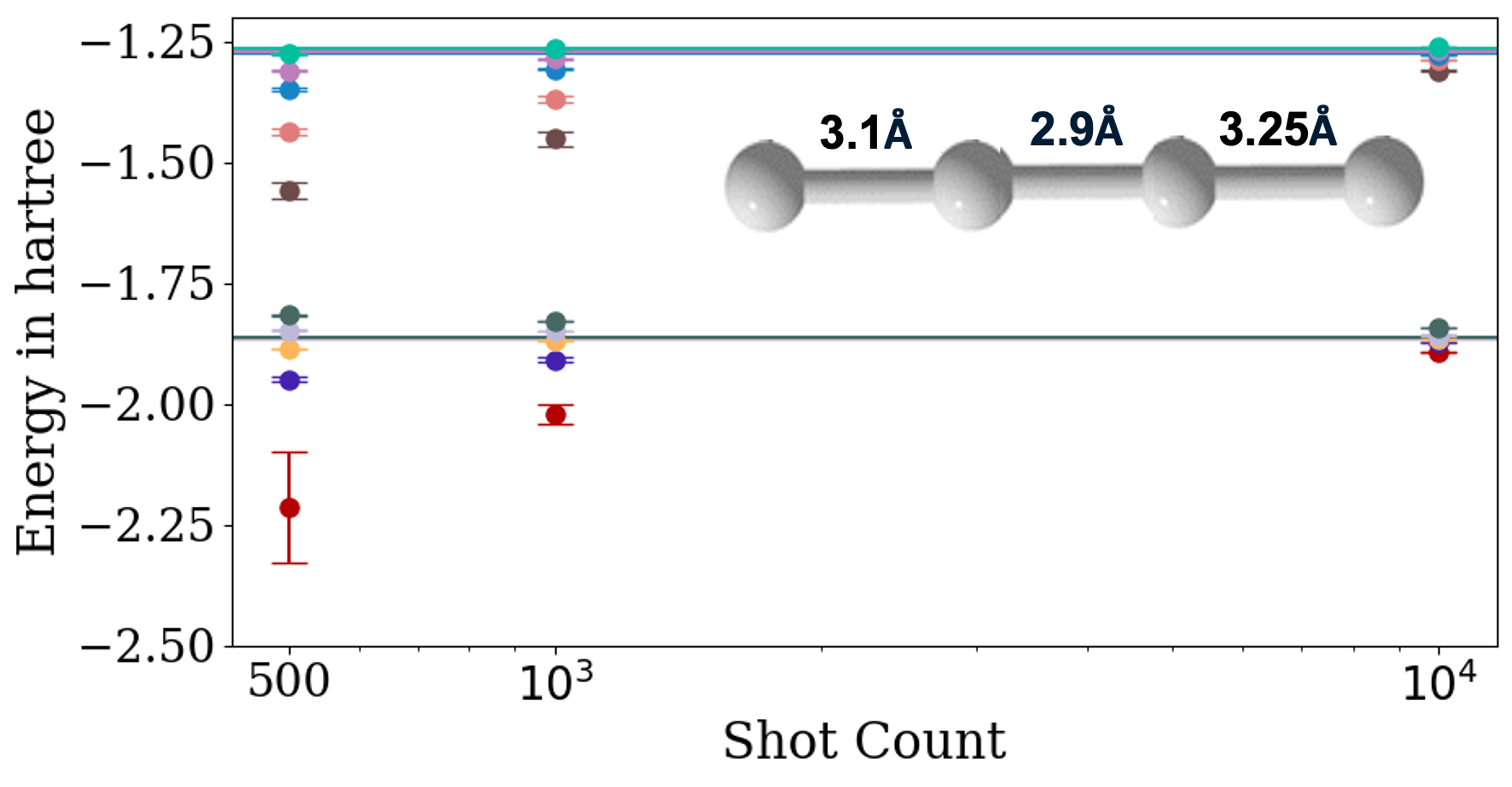}  
        \caption{condition number of 151.9}
        \label{fig:fig2}
    \end{subfigure}
    \label{fig:combined}
 
    \begin{subfigure}[b]{0.45\textwidth}
        \centering
        \includegraphics[width=\textwidth]{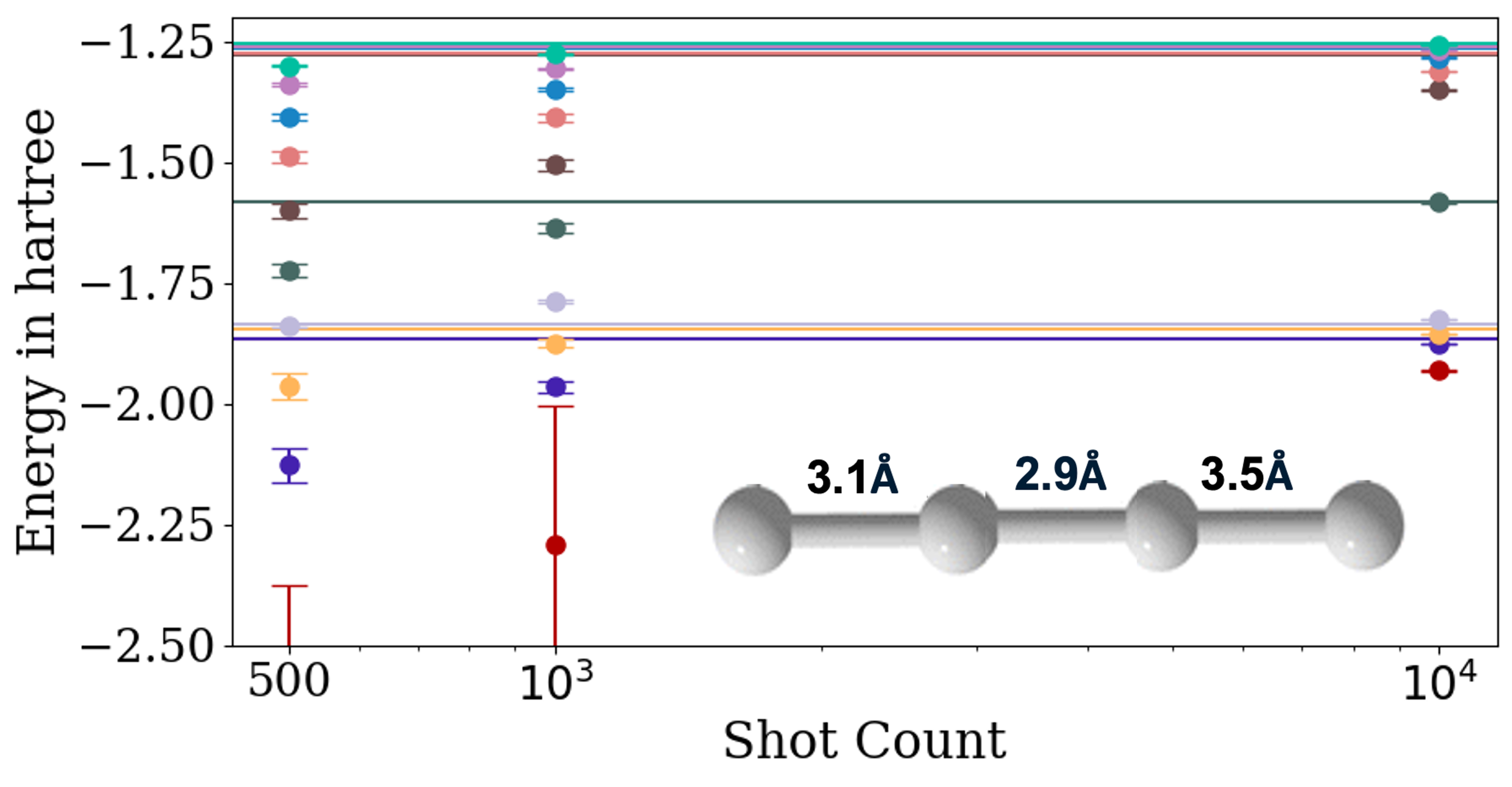}  
        \caption{condition number of 318.8}
        \label{fig:fig3}
    \end{subfigure}
    \hspace{0.5cm}  
    \begin{subfigure}[b]{0.45\textwidth}
        \centering
        \includegraphics[width=\textwidth]{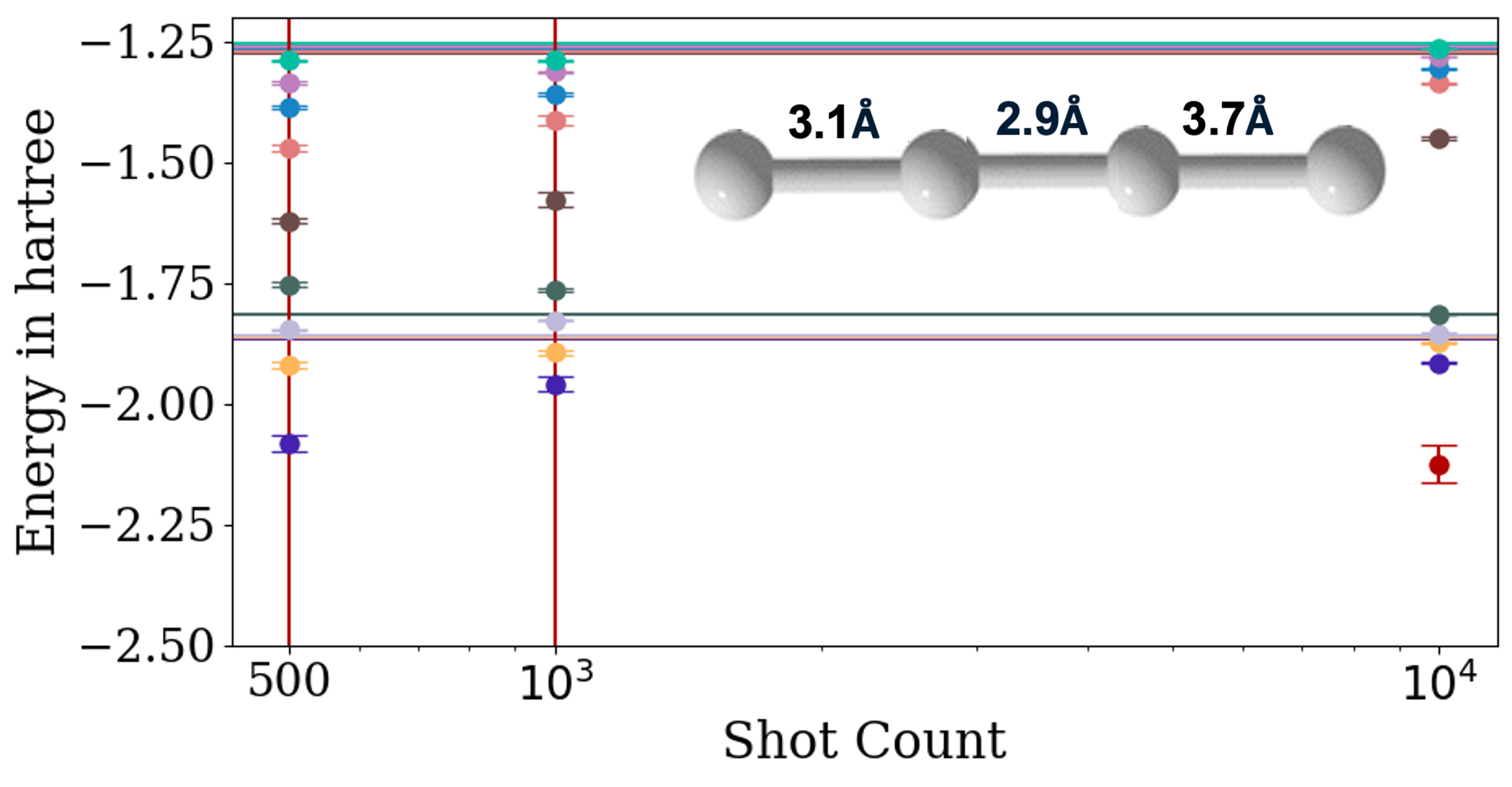}  
        \caption{condition number of 592.5}
        \label{fig:fig4}
    \end{subfigure}
 \caption{Eigenstates computed using QSE in the case of increasing condition number of overlap matrix of QSE (condition number indicated on subfigures). The solid lines represent the exact solution (corresponding to an infinite number of shots) while the spheres and bars represent the mean and variance of the data in 10 simulations using the number of shots per matrix element on the \textit{x}-axis. The geometry of the linear H$_4$ molecule is shown in the subfigures.  This figure highlights the rapid amplification of sampling-noise sensitivity in QSE as $\kappa(\mathbf{S})$ increases (from 121.5 to 592.5), consistent with the $1/\lambda_{\min}(\mathbf{S})$ error-amplification factor in Eq.~(24) and the shot-scaling trend in Eq.~(28).}
    \label{fig:condmed}
\end{figure*}  

\subsection{Medium condition number case}

\begin{figure}[h]
    \centering
    
        \includegraphics[width=0.45\textwidth]{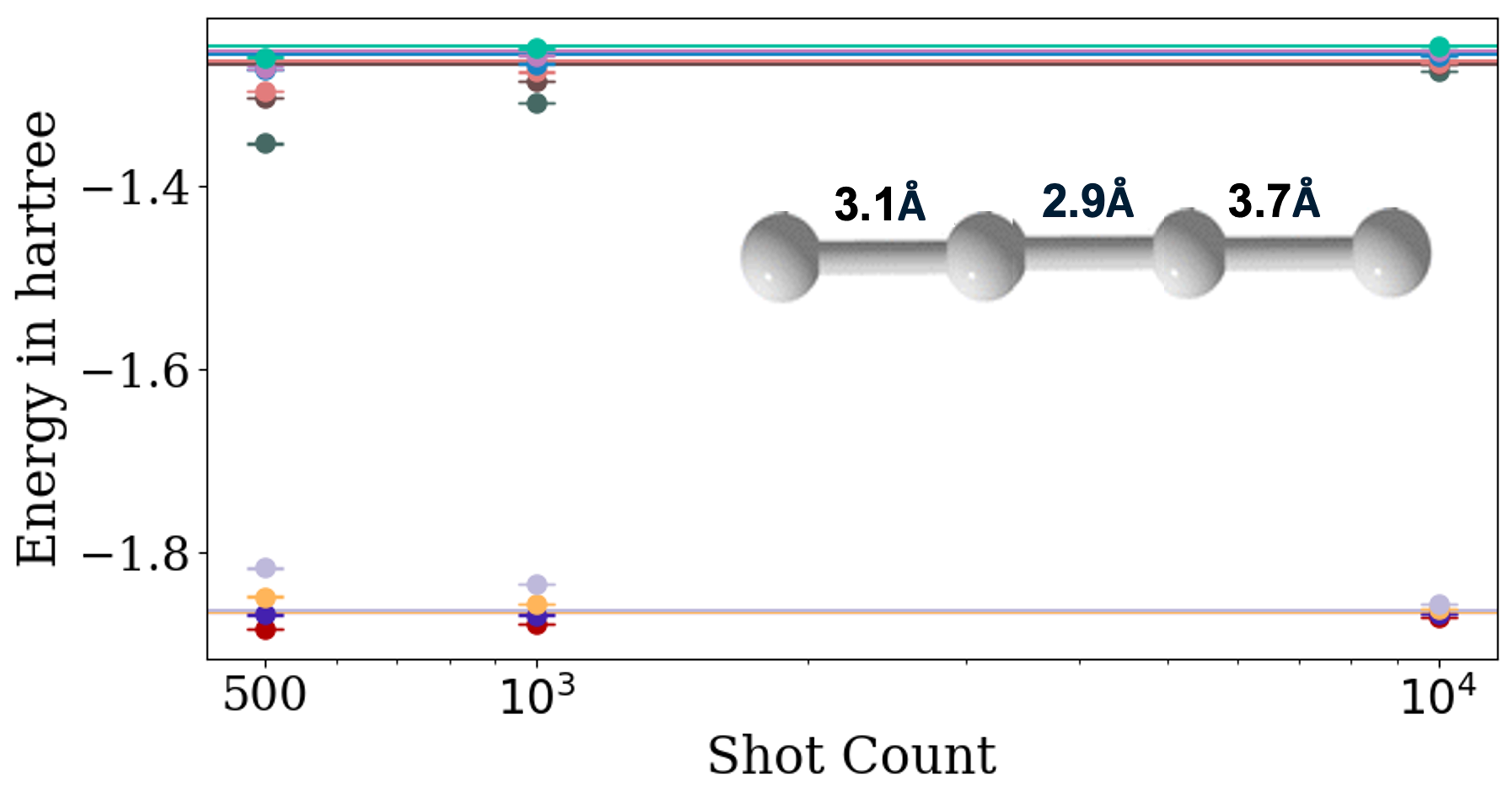}  
 \caption{Eigenstates computed using q-sc-EOM method for the H$_4$ molecular geometry with condition number of overlap matrix in QSE of 592.5. The solid lines, the spheres and the bars have the same meaning as in previous plots.  This figure illustrates that q-sc-EOM remains stable for the same geometry where QSE exhibits a moderately ill-conditioned overlap matrix ($\kappa(\mathbf{S})=592.5$), consistent with the absence of overlap-matrix inversion when $\mathbf{S}=\mathbf{I}$.}
    \label{fig:592qsceom}
\end{figure}

\begin{figure}
    \centering
    \includegraphics[width=0.45\textwidth]{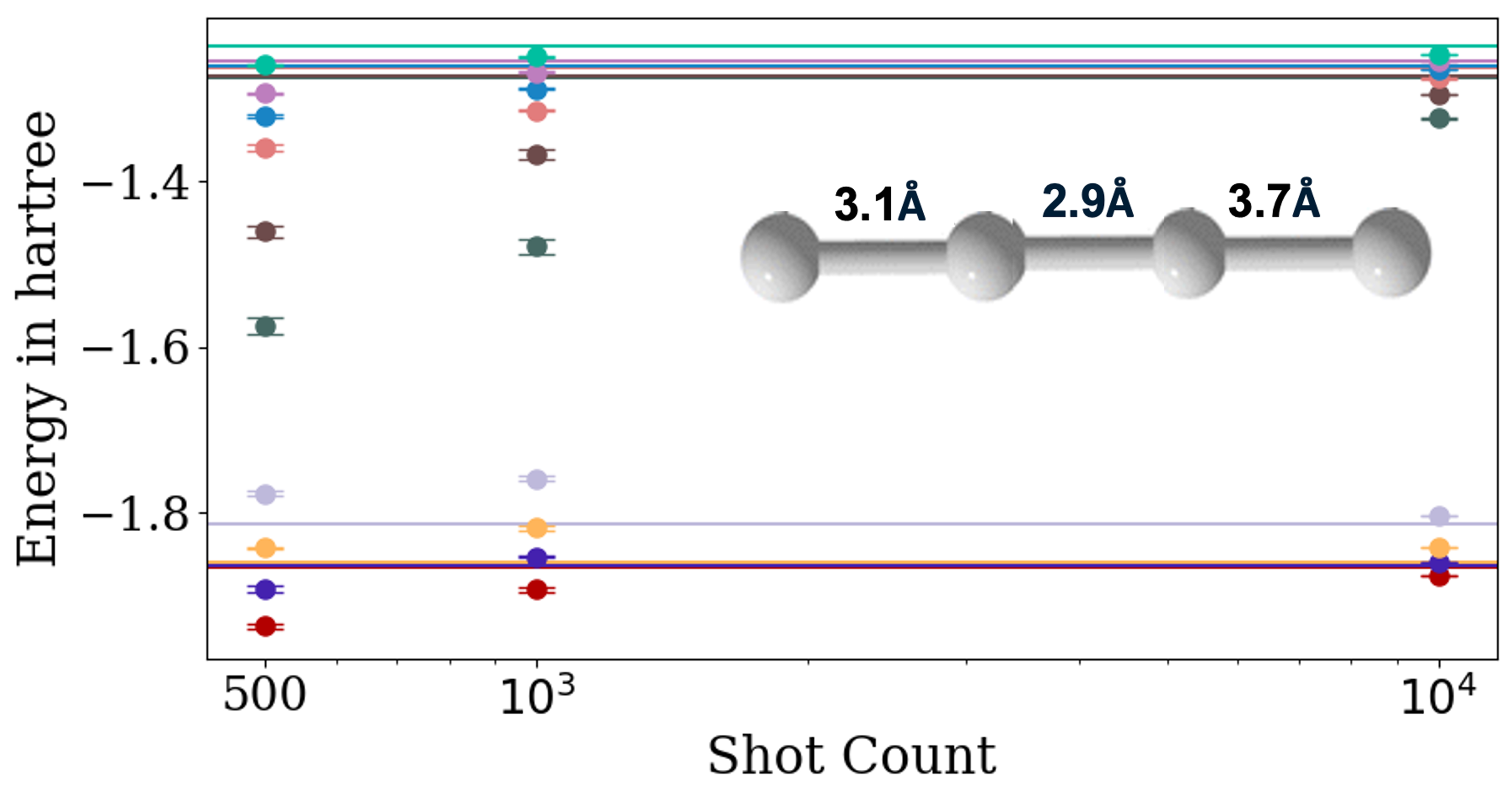} 
    \caption{Eigenstates computed using modified QSE method with artificially exact overlap matrix for the H$_4$ molecular geometry with condition number of overlap matrix in QSE of 592.5. The solid lines, the spheres and the bars have the same meaning as in previous plots.  This figure shows that using an (artificially) exact overlap matrix substantially reduces QSE errors, supporting the interpretation that noise in $\mathbf{S}$ and its subsequent inversion are key contributors to the observed instability.}
    \label{fig:qseexact}
\end{figure}

The comparison of methods becomes much more involved in the case the condition number of the overlap matrix is in the medium range. 
We find that the QSE method generally becomes much more unstable to sampling errors as the condition number of the overlap matrix increases. 
This is consistent with the theoretical argument that the overlap matrix inversion step of solving a generalized eigenvalue equation (see section \ref{GEP}) becomes increasingly sensitive to errors as the condition number of the overlap matrix increases. This phenomenon would also be seen in classical computers, but it would only have a significant impact when the condition number is extremely large, as the errors in classical computers are usually small in the range of 10$^{-10}$ to 10$^{-16}$ when double precision numbers are used.  In quantum computers, these errors can be significant even if only statistical sampling errors are taken into consideration, as done in this study, and therefore, medium range condition number cases are also greatly affected.

Fig. \ref{fig:condmed} shows that errors in the mean and variance in the measured energies increase sharply with the increase in the condition number of the overlap matrix in QSE. 
The four plots examine condition number cases ranging between 120 and 600. 
Fig. \ref{fig:fig1} through \ref{fig:fig4} shows cases of linear H$_4$ molecules with increasing condition numbers of the overlap matrix. The geometry of the molecule is provided in the plots. As the condition number increases from 121.5 to 592.5, a significant increase in errors and variance of the measured eigenvalues is observed. The lowest eigenstate seems to be the worst affected.
 Significant variations of eigenvalues can be seen even when 10,000 shots are used for each matrix element. 
 
 However, as expected, q-sc-EOM is unaffected by this problem and can provide a reasonable solution 
 as demonstrated in Fig. \ref{fig:592qsceom} 
 even for the same linear H$_4$ molecular case where QSE shows a condition number of its overlap matrix to be 592.5.
 This is because q-sc-EOM has an orthonormal set of basis vectors forming the diagonalization space, resulting in the overlap matrix being an identity matrix. This converts the generalized eigenvalue problem to an eigenvalue problem, thus not requiring the noise-sensitive step of inversion of the overlap matrix (see section \ref{GEP}). 
 It should also be noted that if the overlap matrix is artificially made to be exact, these errors are significantly reduced. This can be seen in Fig. \ref{fig:qseexact}, where we have used the exact overlap matrix (estimated at infinite shots). The plot shows that the errors in the eigenstates have reduced significantly, indicating that most of the high errors originate from the error-sensitive step of inversion of the overlap matrix when the condition number of the overlap matrix is high.

\begin{figure}
    \centering
    \includegraphics[width=0.45\textwidth]{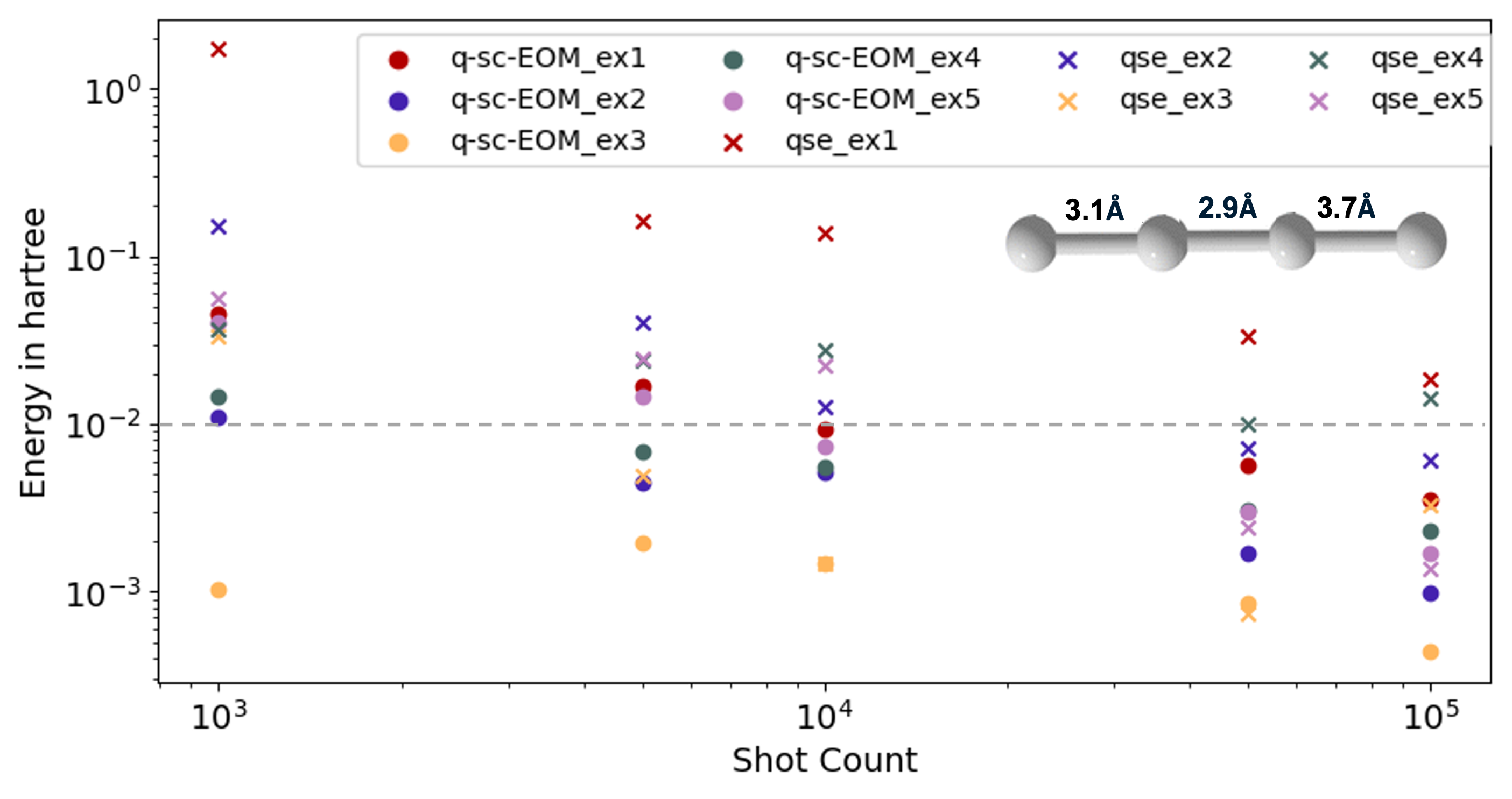} 
    \caption{Error in lowest 5 excited states vs the number of shots for H$_{4}$ molecular geometry with condition number 592.5 (geometry shown in the figure). This figure emphasizes the resource implication of overlap conditioning: for $\kappa(\mathbf{S})=592.5$, QSE requires approximately an order-of-magnitude more shots than q-sc-EOM to achieve comparable accuracy for the lowest excited states, consistent with Eq.~(28).}
    \label{fig:shots}
\end{figure}

\begin{figure*}[t]
    \centering
    \begin{subfigure}[b]{0.45\textwidth}
    \includegraphics[width=\textwidth]{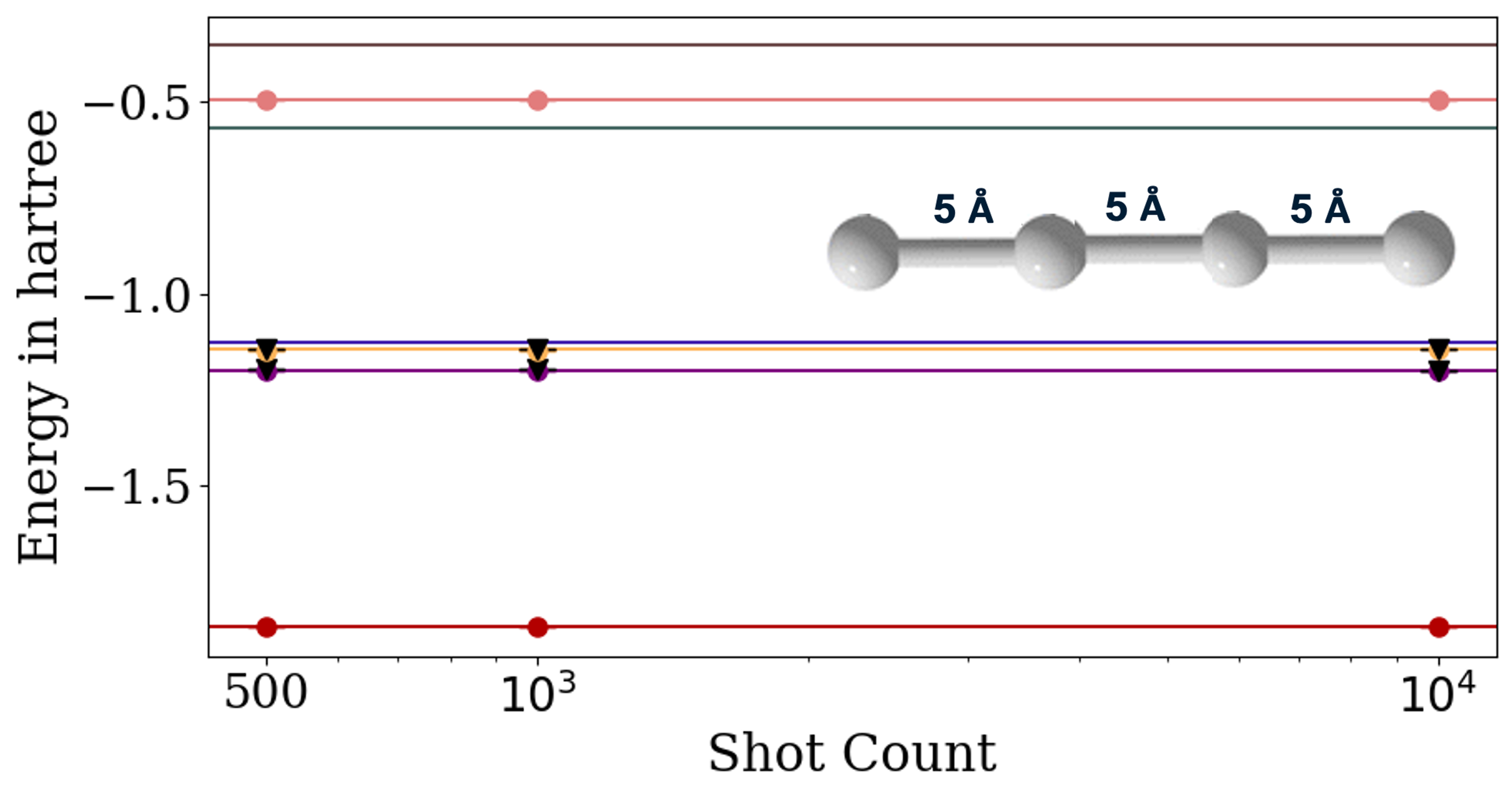} 
    \caption{QSE with thresholding}
    \end{subfigure}
    \begin{subfigure}[b]{0.45\textwidth}
    \includegraphics[width=\textwidth]{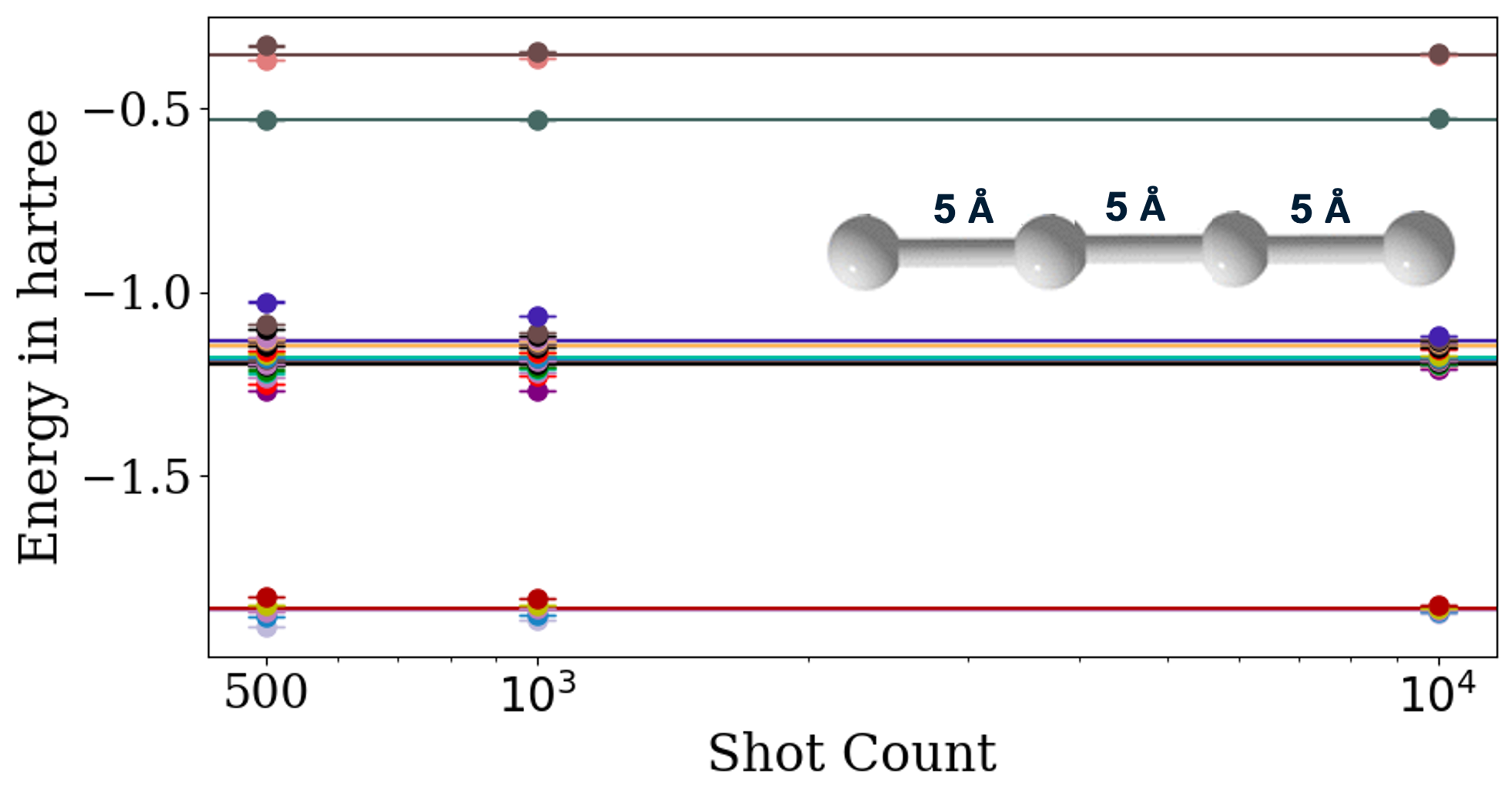} 
    \caption{q-sc-EOM}
\end{subfigure}
\caption{Eigenstates computed using (a) QSE method with thresholding (condition number of overlap matrix is 2.05$\times$10$^{12}$) and (b) q-sc-EOM method (without the need for thresholding) for the linear H$_4$ molecular geometry. The solid lines represent the exact solution (corresponding to an infinite number of shots) while the spheres and bars represent the mean and variance of the data in 10 simulations using the number of shots per matrix element on the \textit{x}-axis. The geometry of the linear H$_4$ molecule is shown in the figure. This figure illustrates that at very large overlap-matrix condition number ($\kappa(\mathbf{S})=2.05\times10^{12}$), QSE requires thresholding to obtain a solvable generalized eigenvalue problem; however, thresholding can compromise spectral completeness (missing/degenerate states) in this example, whereas q-sc-EOM yields a stable solution without thresholding and does not exhibit missing states here.}
\label{fig:threshold}
\end{figure*}

 The error sensitivity of QSE has an important impact on the method's resource requirements. Fig. \ref{fig:shots} shows the number of shots used vs the accuracy of the first five excited states. It can be concluded that for H$_4$ molecule geometry used with condition number of 592.5 of the overlap matrix, 10$^4$ shots per matrix element on q-sc-EOM was able to produce similar accuracy of eigenstates compared with 10$^5$ shots case on QSE. This indicates order-of-magnitude savings of shots in cases with 592.5 condition numbers using q-sc-EOM compared to QSE. This effect will increase as the condition number of the overlap matrix increases.
 Although the effect of sampling errors can be reduced by increasing shot count, other sources of errors may not be completely correctable in this manner.

\subsection{High condition number case: Use of thresholding}


The excited-state energies of linear H$_{4}$ molecules with all bond lengths of 5~$\text{\AA}$ computed using QSE and q-sc-EOM are plotted in Fig.~\ref{fig:threshold}. The condition number of the overlap matrix in QSE in this case is $2.05\times10^{12}$. Due to this very high condition number, the QSE generalized eigenvalue equation could not be solved for a finite number of shots. The standard method to address this problem is thresholding (as discussed in Ref.~\cite{epperly2022theory}). We used a threshold of $5\times10^{-3}$ for $10^3$ and $10^4$ shots, and $1\times10^{-2}$ for 500 shots. These thresholds were selected because they were the smallest values that resolved the singularity in the selected cases. This thresholding process was able to provide solutions to the generalized eigenvalue equation, but the resulting solution has the following problems: \\
(i) (see Fig.~\ref{fig:threshold}) Some excited states are missing from the computed spectrum. These correspond to the green- and brown-colored lines around $-0.3$ to $-0.6$~Hartree and the light-blue line near $-1.1$~Hartree. \\
(ii) Most of the degenerate states in the solution are missing.\\
This is due to the loss of information during thresholding: small eigenvalues (and their associated eigenvectors) of the overlap matrix are removed, which restricts the QSE subspace onto which the Hamiltonian is projected. 
Thresholding should be interpreted not only as a numerical stabilization procedure, but also as an explicit reduction of the effective subspace (model space), which can compromise spectral completeness.
We note that the q-sc-EOM method provides the full set of eigenstates, consistent with the other examples shown in this paper. 
It may appear that QSE with thresholding has higher accuracy at a lower shot count, but this effect may arise because thresholding significantly reduces the matrix size and may be an artifact of solving the generalized eigenvalue equation for much smaller matrices. The accuracy of q-sc-EOM is consistent in all cases studied.

\subsection{Case with NH$_{3}$ system}
\begin{figure*}[t]
    \centering
    
\begin{subfigure}[b]{0.45\textwidth}
    \includegraphics[width=\textwidth]{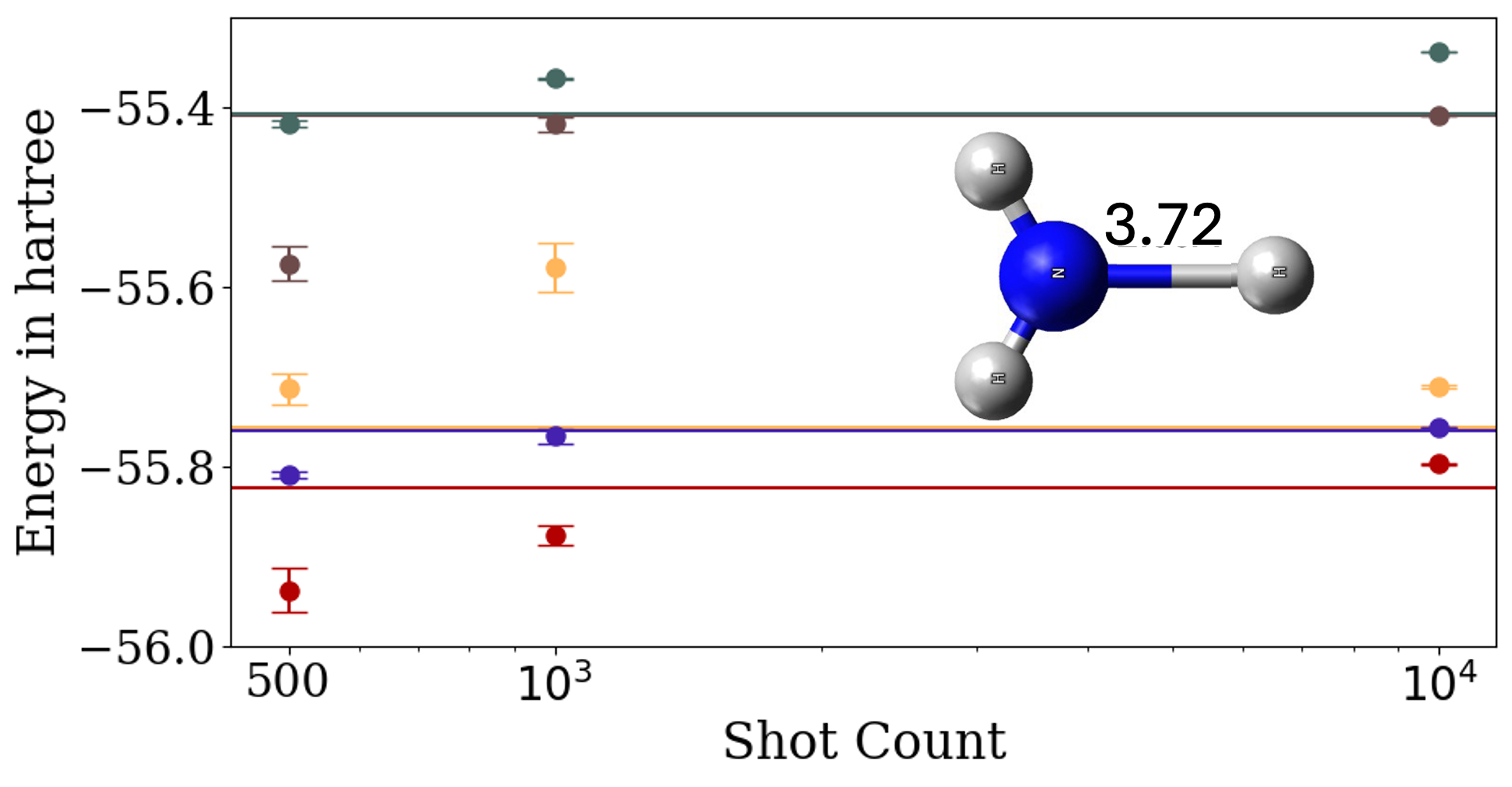}
    \caption{QSE}
    \label{fig:amm1.png}
\end{subfigure}
\begin{subfigure}[b]{0.45\textwidth}
    \includegraphics[width=\textwidth]{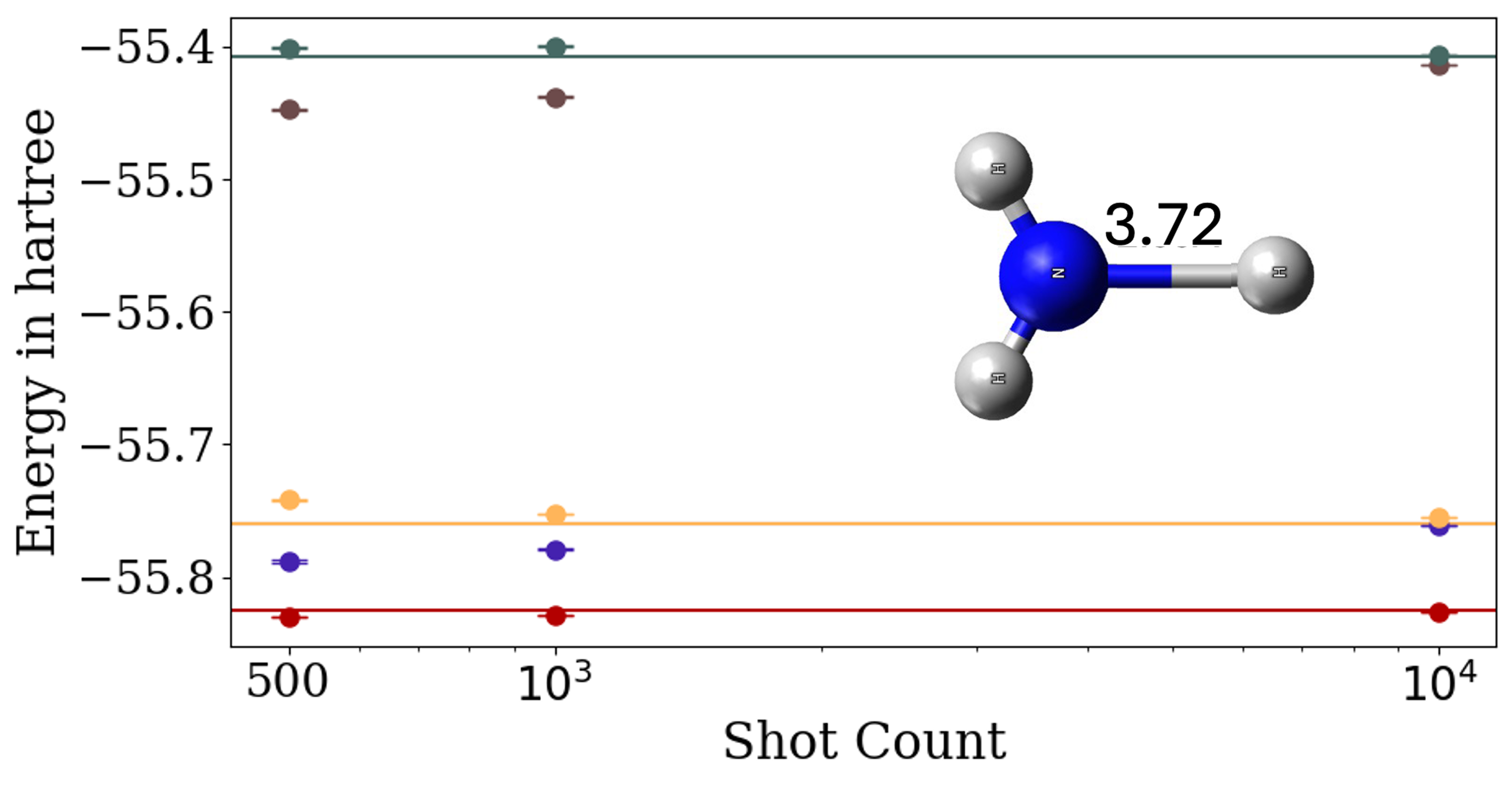} 
    \caption{q-sc-EOM}
    \label{fig:amm2.png}
\end{subfigure}

\caption{Eigenstates of NH$_{3}$ computed using (a) QSE in the case of the condition number of the overlap matrix (condition number is 124.4) and (b) q-sc-EOM. The solid lines represent the exact solution (corresponding to an infinite number of shots) while the spheres and bars represent the mean and variance of the data in 10 simulations using the number of shots per matrix element on the \textit{x}-axis. The geometry of the NH$_{3}$ molecule is shown in the figure. This figure shows that the conditioning-driven instability trends persist in a chemically relevant system: QSE exhibits increased deviations/variance at $\kappa(\mathbf{S})=124.4$, while q-sc-EOM remains comparatively stable.}
\label{fig:amm.png}
\end{figure*}

\begin{figure*}[t]
    \centering
    
\begin{subfigure}[b]{0.45\textwidth}
    \includegraphics[width=\textwidth]{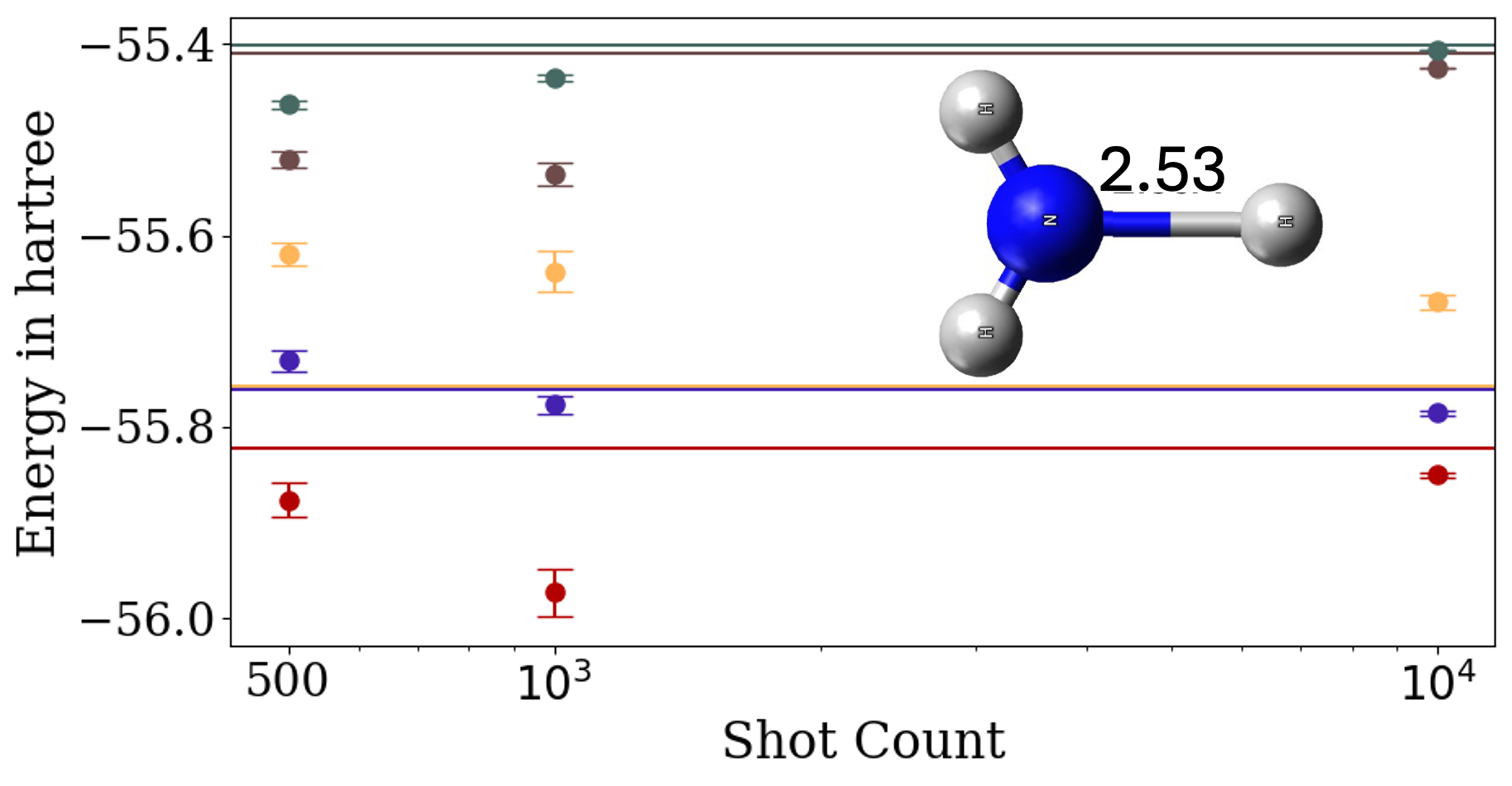}
    \caption{QSE}
    \label{fig:nh1.png}
\end{subfigure}
\begin{subfigure}[b]{0.45\textwidth}
    \includegraphics[width=\textwidth]{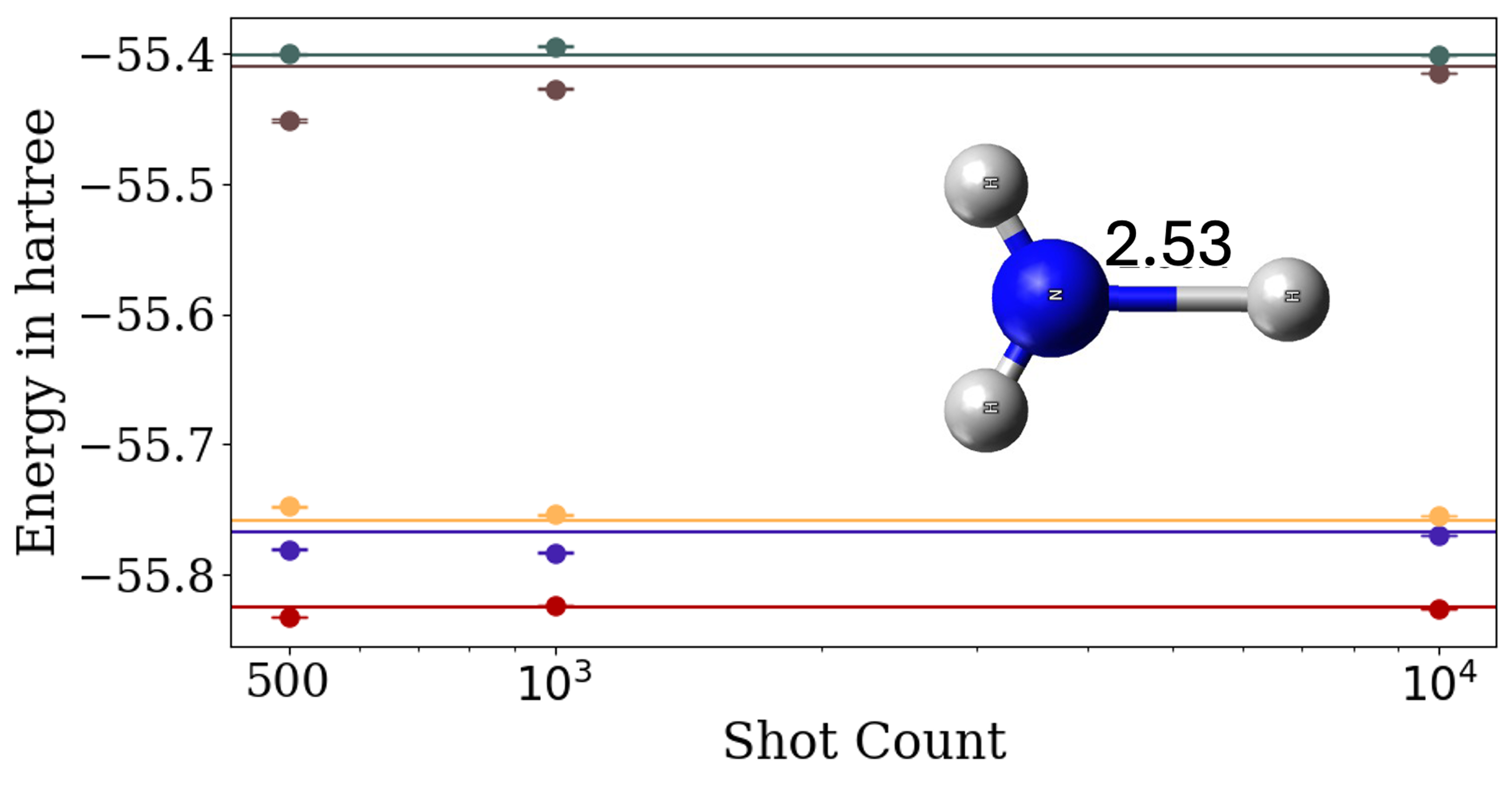} 
    \caption{q-sc-EOM}
    \label{fig:nh2.png}
\end{subfigure}

\caption{Eigenstates of NH$_{3}$ computed using (a) QSE in the case of the condition number of the overlap matrix (condition number is 427.6) and (b) q-sc-EOM. The solid lines represent the exact solution (corresponding to an infinite number of shots) while the spheres and bars represent the mean and variance of the data in 10 simulations using the number of shots per matrix element on the \textit{x}-axis. The geometry of the NH$_{3}$ molecule is shown in the figure. This figure shows that further increase in overlap-matrix conditioning ($\kappa(\mathbf{S})=427.6$) leads to larger QSE instability, while q-sc-EOM remains comparatively stable, highlighting the geometry- and system-dependence of QSE reliability under sampling noise.}
\label{fig:nh.png}
\end{figure*}

Building upon the observations in the H$_{4}$ systems studied above, we extend our analysis to a more chemically relevant molecule, planar NH$_{3}$ as a case study. Using an active space of (4e,4o) and the STO-6G basis set, we observe a similar trend in the excited state energies as seen in the case of H$_{4}$. As the N-H bond of planar NH$_{3}$ is stretched to mimic bond-breaking scenarios, the condition number of the overlap matrix in QSE increased to 124.4 and 427.6. This increase leads to an increase in numerical instability that appears as large deviations in the mean and large standard deviations in the measured energies using QSE, as shown in Fig. \ref{fig:amm1.png} and Fig. \ref{fig:nh1.png}, respectively. It should be noted that in this case, the large amplification of errors due to condition numbers is seen even at relatively small condition numbers compared with the case of H$_4$. This also indicates that the impact of the condition number on numerical instabilities is difficult to predict, as different molecular systems may have different instabilities for states at the same condition number. In contrast, q-sc-EOM remained consistently accurate for both geometries shown in Fig.~\ref{fig:amm2.png} and Fig.~\ref{fig:nh2.png}.

\section{Conclusions}
In this work, we provide both analytical and numerical analyses of the resilience of subspace-based quantum excited state methods to statistical sampling errors on a quantum computer. We choose QSE as a representative method for the class of excited-state methods that require solving the generalized eigenvalue equation on a classical computer with matrix elements evaluated on a quantum computer. We compared this with the q-sc-EOM method, which has an orthonormal set of excited wavefunctions, resulting in a working equation that requires solving an eigenvalue equation instead of the generalized eigenvalue equation.
We divided the possible scenarios into three cases: the low condition number of the overlap matrix in QSE, the medium condition number, and the high condition number. 

We find that the performance of QSE and q-sc-EOM are similar when the condition number of the overlap matrix in QSE is small.
At medium condition numbers, increasingly high errors in eigenstates arise in QSE as the condition number increases. This is because the inversion of an overlap matrix in the presence of noise is error-prone when the condition number is high enough. 
We note that although accuracy can be improved in the case of medium condition number with an increase in shots (if only shot noise is taken into consideration), it will require an order of magnitude more shots to reach a similar accuracy to q-sc-EOM in the examples studied.
At very large condition numbers, the matrix becomes singular, and no solution can be found without the use of the thresholding technique~\cite{epperly2022theory}.
Although the thresholding technique was able to solve the generalized eigenvalue equation, it resulted in missing excited states from the solution.
This could be a potential problem in future chemical studies using these methods.
We show that the dominant instability mechanism associated with solving a generalized eigenvalue problem with an ill-conditioned overlap matrix---in particular, the noise-sensitive inversion of $\mathbf{S}$---can be avoided in q-sc-EOM, whose working equation is a standard eigenvalue problem with an analytically identity overlap matrix ($\mathbf{S}=\mathbf{I}$).
Our analytical study aligns with the numerical findings, and we find that the errors in eigenvalues are magnified in the generalized eigenvalue equation compared with the standard eigenvalue equation. The error magnification depends on the magnitude of the eigenvalue and inversely depends on the minimum eigenvalue of the overlap matrix.

This study highlights the fact that QSE and other diagonalization-based excited-state methods that require the solution of a generalized eigenvalue equation are sensitive to sampling errors on quantum computers, while subspace methods that require a solution to the eigenvalue equation, such as q-sc-EOM, are more resilient alternatives  with respect to the overlap-conditioning-driven error amplification mechanism discussed here.
The conclusions are expected to qualitatively hold for other sources of noise as well, since the condition number of the overlap matrix will magnify the impact of other sources of errors in a similar way (in line with our analytical results). 
Sampling errors are used as a representative example of errors in this study, as we can expect that sampling errors will always exist in Hamiltonian expectation value estimation, even on a perfect quantum computer. 
The condition number of the overlap matrix in methods such as QSE can vary widely depending on molecular geometries; such methods could have significant variation in accuracy depending on the system studied. 
There is no general solution to this problem in QSE that we are aware of in the medium-condition-number regime. Although thresholding can help solve the generalized eigenvalue equation in very high-condition number cases, the problem is not completely solved because the solution achieved through thresholding misses some of the excited states. 
We have used only statistical sampling errors in this study, but we believe that the qualitative conclusions will be valid for other sources of noise in the measurement of matrix elements. 

For balance, we note that q-sc-EOM and related linear-response/EOM approaches also have standard limitations. (i) The accuracy of excitation energies depends on the quality of the underlying ground-state wavefunction. (ii) Truncation of the excitation/operator manifold can limit spectral completeness and quantitative accuracy, particularly for states with strong multi-configurational character. (iii) Unlike ground-state VQE, excited-state solutions obtained from these linear-response/EOM constructions do not generally satisfy a strict variational bound. (iv) The measurement cost of estimating the Hamiltonian projection matrix elements can remain substantial for larger systems, however, it can be reduced substantially by quantum Davidson-type approaches~\cite{kim2023two} with additional circuit depth. Moreover, while $\mathbf{S}=\mathbf{I}$ removes the overlap-inversion amplification mechanism, statistical sampling errors and hardware noise in the Hamiltonian matrix elements can still shift eigenvalues according to standard eigenvalue perturbation behavior.

Spectroscopic applications in molecular sciences require the excited states within a specific energy range to be precisely known. Missing states or states with a high degree of uncertainty can lead to mislabeling or incorrect understanding of the electronic structure of the molecule; therefore, methods with a theoretically exact overlap matrix, such as q-sc-EOM, may be advantageous for such applications when spectral completeness is critical, provided that the ground-state quality and chosen excitation manifold are adequate for the target spectrum.

We also note that similar problems can be expected in other ground and excited state quantum algorithms that require solving a generalized eigenvalue equation in the presence of errors, and this study highlights potential challenges in such methods.
\section{Acknowledgment}
AA would like to thank Nick Mayhall, Diksha Dhawan, Vibin Abraham and Ashutosh Kumar for their useful discussions and their comments on the manuscript. AA, BG and SPS acknowledge startup funding from the University of North Dakota. AA and PFK would like to acknowledge the NSF, award number 2427046 and 2429752, for support. All authors acknowledge Amazon Braket and UND Computational Research Center for computing resources.

\bibliography{main}
\clearpage
\section*{For Table of Contents Only}
\begin{center}
\includegraphics[width=\linewidth]{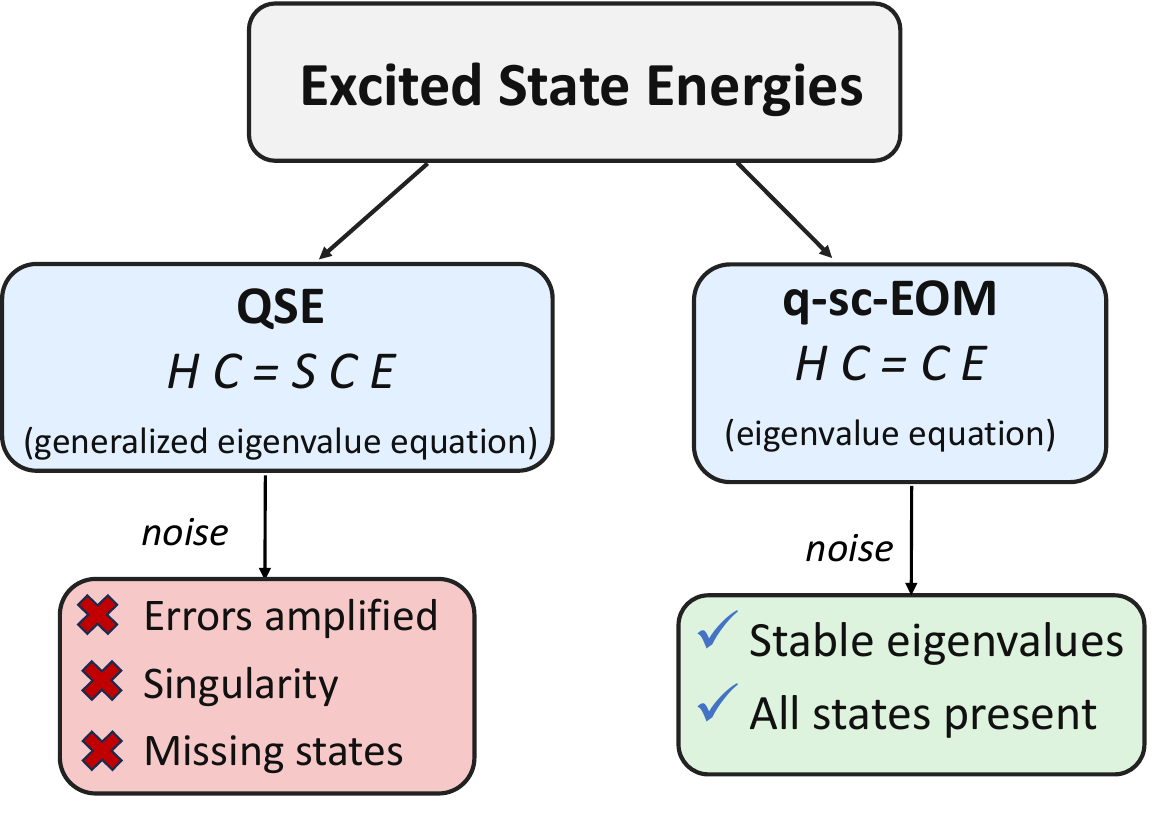}
\end{center}
\end{document}